\documentclass[twocolumn,
               prd,
               aps,
               superscriptaddress,
               tightenlines,
               nofootinbib,
               eqsecnum,
               amsfonts,
               amsmath,
               longbibliography]{revtex4-1}

\usepackage{epsfig}
\usepackage{graphics}
\usepackage{graphicx}
\usepackage[dvipsnames,table]{xcolor}
\usepackage{bm}
\usepackage{natbib}
\usepackage{amssymb}
\usepackage{xspace}
\usepackage[normalem]{ulem} 
\usepackage[colorlinks]{hyperref}
\usepackage[caption=false]{subfig}
\usepackage{url}
\usepackage{float}
\usepackage[bottom]{footmisc}
\usepackage{lineno}
\usepackage{mathrsfs}
\usepackage{makecell}
\usepackage{microtype}
\usepackage{mathtools}
\usepackage{multirow}
\usepackage[inline]{enumitem}
\usepackage{booktabs}
\usepackage{acronym}
\def\newacronym#1#2#3{\gdef#1{\gdef#1{#2\xspace}#3 (#2)\xspace}}
\newacronym{\bhs}{BHs}{black holes}
\newacronym{\bh}{BH}{black hole}
\newacronym{\bbh}{BBH}{binary black-hole}
\newacronym{\ns}{NS}{neutron star}
\newacronym{\nr}{NR}{numerical-relativity}
\newacronym{\eob}{EOB}{effective-one-body}
\newacronym{\gr}{GR}{general relativity}
\newacronym{\gw}{GW}{gravitational-wave}
\newacronym{\snr}{SNR}{signal-to-noise ratio}
\newacronym{\qnm}{QNM}{quasinormal mode}
\newacronym{\pn}{PN}{post-Newtonian}
\newacronym{\imr}{IMR}{inspiral-merger-ringdown}
\newacronym{\rd}{RD}{ringdown}
\newacronym{\LVK}{LVK}{LIGO-Virgo-KAGRA}
\newacronym{\NQC}{NQC}{nonquasicircular}
\newacronym{\ROM}{ROM}{reduced order model}
\newacronym{\DINGO}{DINGO}{Deep Inference for Gravitational-wave Observations}

\AtBeginDocument{%
    \newwrite\bibnotes
    \def\bibnotesext{Notes.bib}
    \immediate\openout\bibnotes=\jobname\bibnotesext
    \immediate\write\bibnotes{@CONTROL{REVTEX41Control}}
    \immediate\write\bibnotes{@CONTROL{%
    apsrev41Control,author="08",editor="1",pages="1",title="0",year="1"}}
    \if@filesw
    \immediate\write\@auxout{\string\citation{apsrev41Control}}%
    \fi
}

\definecolor{aeired}{RGB}{173, 0, 0}
\definecolor{aeiblu}{RGB}{1, 67, 95}
\hypersetup{linkcolor=aeired}
\hypersetup{citecolor=aeiblu}
\hypersetup{urlcolor=aeiblu}

\graphicspath{{plots/}}

\usepackage{perpage}
\MakePerPage{footnote}

\newcommand{\lm}{{\ell m}}
\newcommand{\tm}{t^{\ell m}_{\rm match}}

\newcommand{\df}[1]{\delta f_{\text{#1}}}
\newcommand{\dtau}[1]{\delta \tau_{\text{#1}}}
\newcommand{\fngr}[1]{f_{\text{#1}}}
\newcommand{\taungr}[1]{\tau_{\text{#1}}}

\newcommand{\pseob}{\texttt{pSEOBNRv5PHM}}
\newcommand{\seobfivehm}{{\texttt{SEOBNRv5HM}}}
\newcommand{\seobfivephm}{{{\texttt{SEOBNRv5PHM}}}}

\def\mr{\mathrm}

\def\J2P{J\rightarrow P}
\def\I2J{I\rightarrow J}

\newcommand{\AEI}{\affiliation{Max Planck Institute for Gravitational Physics (Albert Einstein Institute), D-14476 Potsdam, Germany}}
\newcommand{\UMD}{\affiliation{Department of Physics, University of Maryland, College Park, Maryland 20742, USA}}
\newcommand{\UIUC}{\affiliation{Department of Physics and Illinois Center for Advanced Studies of the Universe,\\University of Illinois Urbana-Champaign, Urbana, Illinois 61801, USA}}

\begin{document}

\title{A parametrized spin-precessing inspiral-merger-ringdown\\waveform model for tests of general relativity}

\author{Lorenzo Pompili}
\email{lorenzo.pompili@aei.mpg.de}
\AEI

\author{Elisa Maggio}        \AEI
\author{Hector O. Silva}     \AEI \UIUC
\author{Alessandra Buonanno} \AEI \UMD

\date{\today}

\begin{abstract}
The coalescence of binary black holes (BBHs) provides a unique arena to test general relativity (GR) in the dynamical, strong-field regime.
To this end, we present \pseob, a parametrized, multipolar, spin-precessing waveform model for BBHs in quasicircular orbits, built within the effective-one-body formalism.
Compared to its predecessor, \texttt{pSEOBNRv4HM}, our model introduces parametrized deviations from GR not only in the plunge-merger-ringdown stages, but also in the inspiral phase through modifications to the conservative dynamics. Additionally, it incorporates, for the first time, spin-precession effects.
The free deviation parameters can be used to perform null tests of GR using current and future gravitational-wave observations.
We validate \pseob{} through Bayesian parameter estimation, focusing on the quasinormal-mode frequency and damping time of the $(\ell,m,n) = (2,2,0)$ mode.
Our analysis of synthetic signals from numerical-relativity (NR) simulations of highly precessing BH mergers shows that, while \pseob{} correctly recovers consistency with GR, neglecting spin precession can lead to false detections of deviations from GR even at current detector sensitivity.
Conversely, when analyzing a synthetic signal from a NR simulation of a binary boson-star merger, the model successfully identifies a deviation from a GR BBH signal.
Finally, we reanalyze 12 events from the third Gravitational Wave Transient Catalog.
Using a hierarchical combination of these events, we constrain fractional deviations in the frequency and damping time of the $(2,2,0)$ quasinormal-mode to $\delta {f}_{220}=0.00_{-0.06}^{+0.06}$ and $\delta {\tau}_{220}=0.15_{-0.24}^{+0.26}$ at 90\% credibility.
These results are consistent with those from the LIGO-Virgo-KAGRA Collaboration, which did not account for spin-precession effects.
\end{abstract}

\maketitle


\section{Introduction}
\label{sec:intro}

By the conclusion of the fourth observing run (O4) of the \LVK{} detectors~\cite{LIGOScientific:2014pky, VIRGO:2014yos, KAGRA:2020tym, KAGRA:2021vkt}, it is anticipated that more than 200 \gw{} events, predominantly \bbh{} mergers, will have been observed~\cite{LIGOScientific:2016aoc, KAGRA:2013rdx}.
The coalescence of \bbh{} provides an unparalleled opportunity to test \gr{} in the highly dynamical and strong-field regime~\cite{LIGOScientific:2016lio, LIGOScientific:2019fpa, LIGOScientific:2020tif, LIGOScientific:2021sio}, which is largely inaccessible to other experiments.

To probe \gr{} through \bbh{} events, both theory-specific~\cite{Yunes:2016jcc, Silva:2022srr, Maselli:2023khq, Julie:2024fwy, Maenaut:2024oci} and theory-independent~\cite{Blanchet:1994ez, Arun:2006hn, Yunes:2009ke, Li:2011cg, Agathos:2013upa, Meidam:2017dgf, Brito:2018rfr, Cardoso:2019mqo, Maselli:2019mjd, Ghosh:2021mrv, Mehta:2022pcn, Sanger:2024axs} frameworks have been developed, targeting all stages of binary coalescences: the inspiral, merger, and ringdown.
The ringdown phase, during which the perturbed remnant settles into a Kerr BH by emitting quasinormal modes (QNMs) with discrete frequencies and decay times determined solely by its mass and spin~\cite{Vishveshwara:1970zz,Teukolsky:1973ha,Chandrasekhar:1975zza,Kokkotas:1999bd,Berti:2009kk}, provides a promising testing ground.
Within GR, the ``no-hair'' conjecture \cite{Israel:1967wq,Hawking:1971vc,Carter:1971zc,Robinson:1975bv,Mazur:1982db} posits that an astrophysical, electrically neutral, BH is completely described by its mass and spin only. These two parameters uniquely determine the QNM frequencies.
Therefore, the measurement of at least two QNMs allows to test the consistency between the estimates of mass and spin of the remnant object across multiple frequencies -- this is the fundamental idea behind BH spectroscopy~\cite{Detweiler:1980gk, Dreyer:2003bv, Berti:2005ys, Gossan:2011ha, Meidam:2014jpa, Carullo:2018sfu, Brito:2018rfr, Isi:2019aib, Bhagwat:2019bwv, Maselli:2019mjd, Carullo:2019flw, CalderonBustillo:2020rmh, LIGOScientific:2020tif, LIGOScientific:2021sio, Isi:2021iql, Ghosh:2021mrv, Capano:2021etf, Cotesta:2022pci, Ma:2022wpv, Baibhav:2023clw, Siegel:2024jqd}. Any inconsistency between these estimate
would potentially be indicative of a non-BH nature of the remnant object, or the incompleteness of \gr{} as the underlying theory of gravity.

While the QNM frequencies and decay times depend solely on the remnant properties, the amplitude of each mode and the relative phases between them depend on the properties of the BHs in the binary and the binary dynamics. These quantities can be accurately extracted from \nr{} simulations~\cite{Kamaretsos:2011um, Kamaretsos:2012bs, London:2014cma, London:2018gaq, JimenezForteza:2020cve, Cheung:2023vki, Zhu:2023fnf, Pacilio:2024tdl, MaganaZertuche:2024ajz, Carullo:2024smg}. Leveraging NR-informed amplitude parametrizations, while introducing additional assumptions about the nature of the coalescence, breaks degeneracies between remnant mass, spin, and GR-deviations in the QNM frequencies, that are otherwise present for a single QNM. This approach opens the possibility of measuring deviations from \gr{} even with a single resolvable mode~\cite{Brito:2018rfr,Ghosh:2021mrv, LIGOScientific:2021sio, Gennari:2023gmx}.

A consistent modeling of the merger-ringdown together with the inspiral takes full advantage of the entire signal power, and can further improve the ability to constrain deviations from GR during the ringdown, while also removing dependency on the predicted or estimated start time of the ringdown.
This is the spirit of the parametrized SEOBNR (\texttt{pSEOBNR}) analysis, one of the flagship tests of \gr{} performed within the \LVK{} Collaboration~\cite{LIGOScientific:2020tif, LIGOScientific:2021sio}.
This approach was first introduced in Refs.~\cite{Brito:2018rfr,Ghosh:2021mrv},
that
developed a parametrized model of the ringdown signal as part of the full \imr{} waveforms~\cite{Pan:2011gk,Cotesta:2018fcv} in the \eob{} formalism~\cite{Buonanno:1998gg,Buonanno:2000ef}, calibrated to NR simulations for spinning, nonprecessing BBHs.
The model has been extended to parametrize the plunge-merger stages in Ref.~\cite{Maggio:2022hre}, and to
do
theory-specific tests of GR in the ringdown in Ref.~\cite{Silva:2022srr}.

The \texttt{pSEOBNR} analysis has been applied to perform parametrized ringdown tests on 12 \bbh{} GW signals observed by the \LVK{} Collaboration, showing so far consistency with \gr{} \cite{LIGOScientific:2020tif,LIGOScientific:2021sio}.
The \snr{} of the sources results in measurement errors for the frequency and decay time of the dominant QNM on the order of 10\% and 20\%, respectively, when combining events in a hierarchical way.
Near-future upgrades to the \LVK{} interferometers, and upcoming detectors on the ground, such as Einstein Telescope (ET) and Cosmic Explorer (CE)~\cite{Punturo:2010zz, Evans:2021gyd}, and in space, such as LISA~\cite{LISA:2017pwj}, will significantly increase the number of detected sources. Some events will be observed at \snr{} reaching thousands, enabling tests of GR with unprecedented precision.
This increased sensitivity, however, poses a major challenge for waveform accuracy, as statistical uncertainties approach the systematic biases of the GW approximant models. Several studies predict severe biases in parameter estimation (PE) due to mismodeling errors with the upcoming fifth LVK observational run (O5) and next-generation detectors~\cite{Purrer:2019jcp, Gamba:2020wgg, Dhani:2024jja, Kapil:2024zdn}.

For tests of GR using the full IMR signal, such as the \texttt{pSEOBNR} approach, waveform systematics could lead to false indications of deviations from \gr{}~\cite{Pang:2018hjb, Maggio:2022hre, Hu:2022bji, Toubiana:2023cwr, Gupta:2024gun}.
In Ref.~\cite{Toubiana:2023cwr}, the \texttt{pSEOBNR} analysis was applied to massive BH binaries with LISA, showing that already for SNRs of $\sim 100$, waveform mismodeling could erroneously indicate deviations from GR.
The presence of biases depends strongly on the binary parameters. Merger-dominated signals, where the higher-order modes play a significant role, are especially sensitive to systematic errors~\cite{Toubiana:2023cwr}. For a simulated stellar-mass BBH signal, detectable in O5 with an SNR of 75 and a large fraction of the SNR accumulated in the inspiral phase, the analysis found results consistent with GR~\cite{Ghosh:2021mrv}.
Currently, the \texttt{pSEOBNR} analysis, as well as ringdown-only analyses based on similar templates~\cite{Gennari:2023gmx}, assume BBHs with spins aligned or antialigned with the orbital angular momentum of the binary (i.e., aligned spins).
The analysis of the event GW200129\_065458 in Ref.~\cite{Maggio:2022hre} reported a violation of GR in the merger amplitude of the waveform, while the QNM-deviation parameters remained consistent with GR predictions. This result was interpreted as a false violation of GR, originating either from waveform systematics~\cite{hannam:2021pit} (mismodeling of spin precession that is absent in the baseline GR waveform model) or from data-quality issues~\cite{payne:2022spz}.
Systematic biases in testing GR due to the absence or mismodeling of spin-precession effects have been highlighted in several other studies~\cite{Foo:2024exr, Chandramouli:2024vhw}. These results
underscore the importance of improving waveform models and incorporating all relevant physical effects to perform reliable tests of GR.

In this work, we address two key advancements to improve the reliability and flexibility of the \texttt{pSEOBNR} tests of GR.
First, we extend the \texttt{pSEOBNR} framework to spin-precessing binaries, by introducing a number of free parameters in the \texttt{SEOBNRv5PHM} model~\cite{Pompili:2023tna,Khalil:2023kep,vandeMeent:2023ols,Ramos-Buades:2023ehm}, a state-of-the-art multipolar, spin-precessing EOB waveform model for BBHs in quasicircular orbits.
Additionally, we incorporate deviations across multiple stages of the coalescence. In the ringdown stage, we add fractional deviations to the frequency and damping time of the fundamental QNMs. For the plunge-merger stage, following Ref.~\cite{Maggio:2022hre}, we add fractional deviations to the merger amplitude and frequency of the waveform, as well as the instant at which the amplitude peaks.
Most notably, we introduce -- for the first time -- parametrized corrections in the inspiral stage by modifying the NR-calibration parameters of the model within the EOB Hamiltonian, which describes the conservative dynamics of the binary. This approach is similar in spirit to the \texttt{TIGER} framework~\cite{Agathos:2013upa, Meidam:2017dgf}, which also modifies phenomenological NR-calibration coefficients in the late inspiral stage of \texttt{IMRPhenom} models. Adding extra flexibility to our parametrized model allows for capturing more generic deviations from GR, which are generally expected to affect all stages of the coalescence.
When suitable priors informed by NR uncertainties and fitting errors are employed, corrections to NR-calibration parameters can also be used to marginalize over waveform uncertainties~\cite{Pompili:2024yec, Bachhar:2024olc, Breschi:2022xnc}.
We name the new parametrized waveform model \pseob.\footnote{
\pseob{} is publicly available through the Python package \texttt{pySEOBNR} \cite{Mihaylov:2023bkc}.
A tutorial notebook is available at: \url{https://waveforms.docs.ligo.org/software/pyseobnr}.
Stable versions of \texttt{pySEOBNR} are published through the python package index (PyPI), and can be installed via ~\texttt{pip install pyseobnr}.
}

In this work, while we describe the implementation and morphology of all corrections, we validate our model through Bayesian PE for the ringdown of the $(\ell,m,n)=(2,2,0)$ mode, as it is the primary test currently performed in \LVK{} analyses with \texttt{pSEOBNR} models. The $(2,2,0)$ mode is the dominant QNM in quasicircular mergers, except for highly precessing and asymmetric configurations~\cite{Hughes:2019zmt, Lim:2022veo, Siegel:2023lxl, Zhu:2023fnf}.
By recovering synthetic signals from NR simulations of highly precessing BH mergers~\cite{Boyle:2019kee}, we demonstrate that \pseob{} correctly recovers consistency with GR. In contrast, neglecting spin-precession effects leads to false indications of deviations from GR, even at current detector sensitivity.
Conversely, when analyzing a synthetic signal from a NR simulation of a binary boson-star merger~\cite{Evstafyeva:2024qvp}, the model successfully identifies a deviation from a GR BBH signal.

Finally, we reanalyze 12 events from the third Gravitational-Wave Transient Catalog (GWTC-3) that were previously analyzed using an earlier version of the \texttt{pSEOBNR} model for aligned-spin binaries, \texttt{pSEOBNRv4HM}~\cite{Ghosh:2021mrv}, by the \LVK{} Collaboration~\cite{LIGOScientific:2020tif, LIGOScientific:2021sio}.

The paper is organized as follows. In Sec.~\ref{sec:waveform}, we describe the construction of the \pseob{} model starting
from the baseline \texttt{SEOBNRv5PHM} model, and introduce the non-GR parameters that describe potential deviations from GR during
the different stages of the coalescence. In Sec.~\ref{sec:morpho}, we conduct a detailed study of the morphology of the parametrized waveform to identify which parts of the waveform are affected when the non-GR parameters are varied individually.
In Sec.~\ref{sec:pe_injections}, we perform synthetic-signal injection studies, using \nr{} simulations of binary BHs and boson stars. We apply our parametrized model to real data in Sec.~\ref{sec:pe_real_data}, by reanalyzing 12 events from GWTC-3.
Finally, we summarize our conclusions and future work in Sec.~\ref{sec:conclusions}.

We use geometrical units $G = 1 = c$ unless stated otherwise.


\section{The parametrized waveform model}
\label{sec:waveform}

We start this section with a reminder of the construction of the multipolar spin-precessing \texttt{SEOBNRv5PHM} model~\cite{Ramos-Buades:2023ehm}.
We then describe, in Sec.~\ref{subsec:tgr_parameters}, how we deform this baseline model by introducing deformations away from \gr in the inspiral, merger, and ringdown stages, highlighting differences with respect to earlier iterations of the \texttt{pSEOBNR} models.

We consider a binary with masses $m_1$ and $m_2$, with $m_1 \geq m_2$, and spins $\bm{S}_1$ and $\bm{S}_2$. We define the following combinations of the masses
\begin{equation}
\begin{gathered}
M \equiv m_1 + m_2, \quad \mu \equiv \frac{m_1m_2}{M}, \quad \nu \equiv \frac{\mu}{M},  \\
q \equiv \frac{m_1}{m_2}, \quad \mathcal{M} = \nu^{3/5} M.
\end{gathered}
\end{equation}
We also define the dimensionless spin vectors
\begin{gather}
\bm{\chi}_{\mr i} \equiv \frac{\bm{a}_{\mr i}}{m_{\mr i}} = \frac{\bm{S}_{\mr i}}{m_{\mr i}^2},
\end{gather}
the effective spin parameter $\chi_{\rm eff}$ \cite{Racine:2008qv,Santamaria:2010yb},
\begin{equation}
\chi_{\rm eff} = \frac{m_1\chi^\parallel_1+m_2\chi^\parallel_2}{m_1+m_2},
\label{eq:chi_eff}
\end{equation}
and the effective precessing-spin parameter $\chi_{\rm p}$ \cite{Schmidt:2014iyl},
\begin{equation}
\chi_{\rm p} = \frac{1}{B_1 m_1^2}\max \left( B_1 m_1^2 \chi_{1}^\perp, B_2 m_2^2 \chi_{2}^\perp \right),
\label{eq:chi_p}
\end{equation}
where $B_1 = 2+3m_2/(2 m_1)$, $B_2 = 2+3m_1/(2 m_2)$, and we have divided $\bm{\chi}_i$ into its aligned-spin component $(\chi^\parallel_i)$ and in-plane component $(\chi^\perp_i)$.

\subsection{Overview of the \texttt{SEOBNRv5PHM} waveform model}

\subsubsection{Effective-one-body dynamics of precessing-spin binary black holes}

In the EOB formalism~\cite{Buonanno:1998gg, Buonanno:2000ef, Damour:2000we, Damour:2001tu, Buonanno:2005xu}, the two-body conservative dynamics is described by an Hamiltonian $H_{\rm EOB}$,
\begin{equation}
H_{\rm EOB} = M \sqrt{1+2 \nu \left(\frac{H_{\rm eff}}{\mu}-1 \right)}\,,
\end{equation}
where the effective Hamiltonian $H_{\rm eff}$ describes a test mass $\mu$ moving in a deformed Kerr spacetime of mass $M$, the deformation parameter being the symmetric mass ratio $0 \leq \nu \leq 1/4$.

A common strategy for building precessing-spin waveforms is to start from aligned-spin waveforms in the so-called \textit{coprecessing frame}, in which the $z$-axis remains perpendicular to the instantaneous orbital plane, and then applying a suitable rotation to the inertial frame of the observer~\cite{Buonanno:2002fy,Schmidt:2010it,Boyle:2011gg,OShaughnessy:2011pmr,Schmidt:2012rh}.
More specifically, \texttt{SEOBNRv5PHM} builds on previous studies that employed aligned-spin orbital dynamics in the coprecessing frame coupled to post-Newtonian (PN) expanded precessing-spin equations~\cite{Estelles:2020twz,Akcay:2020qrj,Gamba:2021ydi}, to mitigate the computational expense of solving the equations of motion using the full precessing-spin EOB Hamiltonian~\cite{Pan:2013rra, Babak:2016tgq, Ossokine:2020kjp}.

To model precessional effects more accurately, \texttt{SEOBNRv5PHM} extends beyond this approach by using, in the coprecessing frame, an EOB Hamiltonian that includes partial precessional effects ($H_{\rm EOB}^\text{pprec}$), in the form of orbit-averaged in-plane spin contributions for circular orbits~\cite{Khalil:2023kep}. The Hamiltonian $H_{\rm EOB}^\text{pprec}$ reduces in the aligned-spin limit to the one used in the \seobfivehm{} model~\cite{Pompili:2023tna}.
This Hamiltonian features two higher-order PN terms ($a_6$, $d_{\rm SO}$) which are calibrated to 442 aligned-spin NR waveforms from the Simulating eXtreme Spacetimes (SXS) Collaboration~\cite{Boyle:2019kee}. The parameter $a_6$ is a 5PN nonspinning coefficient and $d_{\rm SO}$ is a 4.5PN spin-orbit coefficient in $H_{\rm{EOB}}$.

In the \texttt{SEOBNRv5PHM} model, the equations of motion in the coprecessing frame have the same form of the evolution equations
for aligned-spin binaries, and read~\cite{Ramos-Buades:2023ehm}:
\begin{equation}
\label{eq:EOBEOMs}
\begin{aligned}
\dot{r}&=\xi(r) \frac{\partial H_{\rm EOB}^\text{pprec}}{\partial p_{r_*}}, \quad
&\dot{\phi} &=\frac{\partial H_{\rm EOB}^\text{pprec}}{\partial p_{\phi}},\\
\dot{p}_{r_*}&=-\xi(r)\frac{\partial H_{\rm EOB}^\text{pprec}}{\partial r} +\mathcal{F}_{r}, \quad
&\dot{p}_{\phi}&=\mathcal{F}_{\phi},
\end{aligned}
\end{equation}
where the dot indicates a time derivative. The evolution of the radial momentum is performed using the tortoise-coordinate $p_{r_*} = p_r \, \xi(r)$, with $\xi(r) = {\rm d} r/{\rm d}r_*$.

The radiation-reaction force is computed as~\cite{Buonanno:2005xu}
\begin{equation}
\mathcal{F_\phi}= -\frac{\Phi_E}{\Omega}, \qquad \mathcal{F}_r = \mathcal{F}_\phi \frac{p_{r}}{p_\phi},
\label{eq:RRforceAS}
\end{equation}
where $\Omega \equiv \dot{\phi}$ is the orbital frequency, and $\Phi_E$ is the energy flux radiated by the binary for quasicircular orbits. The energy flux is computed by summing over the contributions of the factorized PN modes (augmented with gravitational-self-force information~\cite{vandeMeent:2023ols}) $h^{\mathrm{F}}_{\ell m}$~\cite{Damour:2007xr,Damour:2007yf,Damour:2008gu,Pan:2010hz},
\begin{equation}
\Phi_E = \frac{\Omega^2}{16 \pi} \sum_{\ell=2}^8 \sum_{m=-\ell}^\ell m^2 |d_L h^{\mathrm{F}}_{\ell m}|^2,
\label{eq:fluxAS}
\end{equation}
where $d_L$ is the luminosity distance from the binary to the observer.

For the evolution of the spins and angular momentum, \seobfivephm{} employs PN-expanded evolution equations that include higher-order PN information and use a spin-supplementary condition consistent with the Hamiltonian $H_{\rm EOB}^\text{pprec}$~\cite{Khalil:2023kep}.
The spin and angular momentum evolution equations are used both for constructing the rotation between different reference frames during the inspiral, as summarized below, and for augmenting the coprecessing frame orbital dynamics, by using, into the Hamiltonian and aligned-spin modes, suitable projections of the spins onto the angular momentum at every time step of the evolution~\cite{Ramos-Buades:2023ehm}.

\subsubsection{Inspiral-plunge waveforms}
\label{sec:InspiralWaveforms}

The gravitational polarizations can be written as
\begin{equation}
    h_+ - i h_\times = \sum_{\ell,m} {}_{-2} Y_{\ell m} (\varphi, \iota) h_{\ell m} (t),
\end{equation}
where ${}_{-2} Y_{\ell m} (\varphi, \iota)$ are the $-2$ spin-weighted spherical harmonics, with $\varphi$ and $\iota$ being the azimuthal and polar angles to the observer. In the EOB framework~\cite{Buonanno:2000ef}, the GW modes in the coprecessing frame are decomposed as
\begin{equation}
\begin{aligned}
h_{\ell m}(t) &=  h_{\ell m}(t)^\mathrm{insp-plunge}\,\theta(t_\mathrm{match}^{\ell m} - t) \\
&\quad + h_{\ell m}(t)^\mathrm{merger-RD}\,\theta(t-t_\mathrm{match}^{\ell m})\,,
\label{eq:EOBGW}
\end{aligned}
\end{equation}
where $\theta(t)$ is the Heaviside step function, $h_{\ell m}^\mathrm{insp-plunge}$ corresponds to the inspiral-plunge part of the waveform, while $h_{\ell m}^\mathrm{merger-RD}$ represents the merger-ringdown waveform.
The matching time $t_\mathrm{match}^{\ell m}$ is chosen to be the peak of the $(2,2)$ mode amplitude for all $(\ell,m)$ modes except $(5,5)$, for which it is taken as the peak of the $(2,2)$ harmonic minus $10 M$~\cite{Pompili:2023tna}.

The peak-time of the $(2,2)$ mode ($t_{\text {peak}}^{22}$) is also calibrated to NR simulations. It is defined as
\begin{equation}
    \label{eq:tISCO}
    t_{\text {peak}}^{22}=t_{\mathrm{ISCO}}+\Delta t_{\mathrm{NR}},
\end{equation}
where $t_{\rm{ISCO}}$ is the time at which $r = r_{\rm{ISCO}}$,
and $r_{\rm{ISCO}}$ is the radius of the innermost-stable circular orbit (ISCO)~\cite{Bardeen:1972fi} of a Kerr BH with the mass and spin of remnant~\cite{Jimenez-Forteza:2016oae, Hofmann:2016yih}.
The parameter $\Delta t_{\rm NR}$ is a free parameter calibrated to aligned-spin NR waveforms~\cite{Pompili:2023tna}.
Note that this choice is different to the one employed in the \texttt{SEOBNRv4} waveform models~\cite{Bohe:2016gbl, Cotesta:2018fcv, Ossokine:2020kjp}, which related the peak-time of the $(2,2)$ mode to the peak of the orbital frequency, rather than the ISCO radius.

The inspiral-plunge modes in the coprecessing frame use the factorized, resummed expressions of the PN GW modes~\cite{Damour:2008gu, Damour:2007xr, Pan:2010hz, Damour:2007yf}, with time-dependent projections of the spins, evaluated on the dynamics obtained from the EOB equations of motion \cite{Buonanno:2000ef, Pan:2011gk}. Their accuracy during the plunge, when the radial motion dominates the dynamics, is further improved by applying the numerically tuned nonquasicircular (NQC) corrections $N_{\ell m}$~\cite{Damour:2007xr},
\begin{equation}
    \label{eq:hlm_insp-plunge}
    h_{\ell m}(t)^\mathrm{insp-plunge} = h^{\mathrm{F}}_{\ell m} N_{\ell m},
\end{equation}
which also allow for a smooth connection between the inspiral-plunge and merger-ringdown waveforms.
Notably, the NQC corrections ensure that the amplitude of the EOB modes $\vert h_{\lm}^{\rm insp-plunge} \vert(\tm)$ and its first two derivatives, and the frequency of the EOB modes $\omega_{\lm}^{\rm insp-plunge}(\tm)$ and its first derivative, match those of the \nr modes of at $t = \tm$ (the so-called \textit{input values}, $|h_{\lm}^{\rm NR}|$ and $\omega_{\lm}^{\rm NR}$). Parameter-space fits for the NR input values are provided in Appendix C of Ref.~\cite{Pompili:2023tna}.

Following Ref.~\cite{Pompili:2023tna}, \seobfivephm{} includes the following modes in the coprecessing frame,
\begin{align}
    \label{eq:modes_v5hm}
    (\ell, |m|) = \{ (2, 2), (2, 1), (3, 3), (3, 2), (4, 4), (4, 3), (5, 5) \} \,.
    \nonumber \\
\end{align}
Negative-$m$ modes are derived from the positive-$m$ ones using the reflection symmetry $h_{\ell m}=(-1)^{\ell} h_{\ell-m}^*$, which is exact for aligned-spin binaries but not for precessing-spin binaries~\cite{Boyle:2014ioa}, even in the coprecessing frame.
\footnote{In parallel to this work, a NR-calibrated model for the antisymmetric mode contributions in the coprecessing frame was developed for \texttt{SEOBNRv5PHM}~\cite{estelles_2025}. While incorporating these modes into the parametrized model is straightforward, we do not include them here, as the version of \texttt{pSEOBNRv5PHM} reviewed within the \LVK{} Collaboration -- intended for use in the O4a analyses -- does not yet include these modes. }

The GW polarizations in the inertial frame of the observer are those required for data-analysis applications.
The \seobfivephm~model makes use of three reference frames (see Fig.~1 of Ref.~\cite{Ramos-Buades:2023ehm}):
\begin{enumerate}
    \item The inertial frame of the observer (\textit{source frame}). Quantities in this frame are denoted with the superscript $\mathrm{I}$.
    \item An inertial frame where the $z$-axis is aligned with the final angular momentum of the system (\textit{$\bm{J}_{\rm f}$-frame}). Quantities in this frame are denoted with the superscript $\mathrm{J}$. This frame facilitates the construction of the merger-ringdown modes. The QNMs are defined with respect to the direction of the final spin, and thus, the description of the ringdown signal as a linear combination of QNMs, is formally valid in this frame.
    \item A noninertial frame which tracks the instantaneous motion of the orbital plane (the \textit{coprecessing frame}). Quantities in this frame are denoted with the superscript $\mathrm{CP}$.
\end{enumerate}

The inertial-frame modes are related to the coprecessing-frame modes by a time-dependent rotation from the coprecessing frame to the $\bm{J}_{\rm f}$-frame, followed by a time-independent rotation from the $\bm{J}_{\rm f}$-frame to the final inertial frame
\begin{equation}
h^{\rm I}_{\ell m}(t)=\sum_{m', m''} \Big(\textbf{R}^{\rm J \rightarrow I}\Big)_{m,m'}\Big(\textbf{R}^{\rm CP \rightarrow J}\Big)_{m',m''}h^{\rm CP}_{\ell m''}(t),
\label{eq:prec_rotations}
\end{equation}
where $\textbf{R}^{\rm X \rightarrow Y}$ denotes the rotation operator from the frame $X$ to the frame $Y$, and $m', m''$ are summation indices over the modes available in the coprecessing frame.
These rotations are implemented in \texttt{pySEOBNR} using quaternions, but can also be expressed in terms of Euler angles $\{\alpha(t),\beta(t),\gamma(t) \}$~\cite{Boyle:2011gg}. Notably, spin precession induces mixing of modes with the same $\ell$ but different $m$, which can reorder the amplitudes of different modes as compared to the nonprecessing scenario \cite{Schmidt:2010it, Boyle:2013nka}, leading to particularly large amplitudes for modes with $\ell=m\neq 2$~\cite{Hughes:2019zmt, Siegel:2023lxl, Zhu:2023fnf}.

\subsubsection{Merger-ringdown waveforms}
\label{sec:MergerWaveforms}
After the merger, the EOB formalism models the transition to the ringdown stage using a phenomenological model~\cite{Damour:2014yha,Pompili:2023tna} based on the QNMs of the remnant BH.
In \seobfivephm, the attachment of the merger-ringdown waveform is performed in the coprecessing frame, using the merger-ringdown multipolar model developed for nonprecessing BBHs in Ref.~\cite{Pompili:2023tna}.

For all harmonics, except for $(\ell, |m|)= (3,2)$ and $(4,3)$ which exhibit postmerger oscillations due to mode mixing~\cite{Buonanno:2006ui, Kelly:2012nd}, the merger-ringdown waveform employs the following \textit{Ansatz}~\citep{Damour:2014yha, Pompili:2023tna},
\begin{align}
\label{RD}
h_{\ell m}^{\textrm{merger-RD}}(t) = \nu \ \tilde{A}_{\ell m}(t)\ e^{i \tilde{\phi}_{\ell m}(t)} \ e^{-i \sigma_{\ell m 0}^{\rm CP}(t-t_{\textrm{match}}^{\ell m})},
\nonumber \\
\end{align}
where $\sigma_{\ell m 0}^{\rm CP} = \sigma_{\ell m}^{\rm R,\,CP} - i \sigma_{\ell m}^{\rm I,\,CP}$ is the complex frequency of the least-damped QNM of the remnant BH, in the coprecessing frame. NR fits are employed to compute the final mass~\cite{Jimenez-Forteza:2016oae} and final spin~\cite{Hofmann:2016yih} of the remnant.
The real and imaginary parts of $\sigma_{\ell m 0}$, whether in the coprecessing or $\bm{J}_{\rm f}$-frame, are related to QNM oscillation frequency and damping time as follows:
\begin{subequations}
\begin{align}
    f_{\lm 0} &= \frac{1}{2\pi} {\rm Re}(\sigma_{\lm 0}) \equiv \frac{1}{2\pi} \omega^{\rm QNM}_{\ell m 0} \,,
    \\
    \tau_{\lm 0} &= - \frac{1}{{\rm Im}(\sigma_{\lm 0})} \,.
\label{eq:qnm_def}
\end{align}
\end{subequations}

The functions $\tilde{A}_{\ell m}$ and $\tilde{\phi}_{\ell m}$ in Eq.~\eqref{RD} are time-dependent amplitude and phase functions (see Sec.~III of Ref.~\cite{Pompili:2023tna} for explicit expressions). Even though source-driven effects, overtones, and nonlinearities are not explicitly included in Eq.~\eqref{RD}, these effects are effectively included in the merger-ringdown model, as the phenomenological \textit{Ansätze} are calibrated against NR simulations.

To account for mode-mixing in the $(3, 2)$ and $(4, 3)$ modes, the same construction is applied to the corresponding \textit{spheroidal harmonics}~\cite{Berti:2005gp} $(3, 2, 0)$ and $(4, 3, 0)$, which maintain a monotonic amplitude and frequency evolution. The $(3, 2)$ and $(4, 3)$ spherical harmonics can be reconstructed by combining the $(3, 2, 0)$ and $(4, 3, 0)$ spheroidal harmonics with the $(2, 2)$ and $(3, 3)$ spherical harmonics, using appropriate mode-mixing coefficients~\cite{Berti:2014fga}.

The calculation of the waveform in the inertial observer's frame requires a description of the coprecessing frame Euler angles $\{\alpha(t),\beta(t),\gamma(t) \}$ which extends beyond merger.
\texttt{SEOBNRv5PHM} makes use of a phenomenological prescription based on insights from NR simulations~\cite{OShaughnessy:2012iol}. Specifically, it has been shown that the coprecessing frame continues to precess approximately around the direction of the final angular momentum with a precession frequency, $\omega_{\rm prec}$, proportional to the difference between the lowest overtone of the $(2,2,0)$ and $(2,1,0)$ QNM frequencies. This phenomenology leads to the following expressions for the merger-ringdown angles relating the $\bm{J}_{\rm f}$-frame and the coprecessing frame used in \seobfivephm,
\begin{align}
    \label{eq:euler_RD}
    \alpha^{\rm merger-RD} &= \alpha(t_{\rm match}) + \omega_{\rm prec}(t-t_{\rm match}), \nonumber \\
    \beta^{\rm merger-RD} &= \beta(t_{\rm match}),\\
    \gamma^{\rm merger-RD} &= \gamma(t_{\rm match}) - \omega_{\rm prec}(t-t_{\rm match}) \cos \beta(t_{\rm match}),
    \nonumber
\end{align}
where $t_{\rm match}=t_{\text {peak}}^{22}$ is the matching time of the merger-ringdown model.
This rotation prescription provides an accurate approximation if the $(2,0,0)$ QNM amplitude in the $\bm{J}_{\rm f}$-frame is negligible. However, for highly precessing binaries the $(2,0,0)$ QNM can be strongly excited, and this assumption may be a source of systematics~\cite{Zhu:2023fnf}.

The behavior studied in Ref.~\cite{OShaughnessy:2012iol} describes prograde-spin configurations, where the remnant spin is positively aligned with the orbital angular momentum at merger. Following Ref.~\cite{Ossokine:2020kjp}, \seobfivephm{} extends the prescription to retrograde-spin cases by imposing simple precession around the final spin at a rate $\omega_{\rm prec} \geq 0$
\begin{equation}
    \label{eq:omega_prec}
    \omega_{\rm prec} =
    \begin{cases}
        \, \omega_{220}^{\rm QNM,\,J}(\chi_f) - \omega_{210}^{\rm QNM,\,J}(\chi_f)   \quad  \bm{\chi}_f \cdot \bm{L}_f > 0 \\ \\
        \, \omega_{210}^{\rm QNM,\,J}(\chi_f) - \omega_{220}^{\rm QNM,\,J}(\chi_f)   \quad  \bm{\chi}_f \cdot \bm{L}_f < 0
    \end{cases} \,,
\end{equation}
that depends on whether the total angular momentum at merger $\bm{\chi}_f \propto {\bm J}_f$ is aligned or not with the orbital angular momentum at merger $\bm{L}_f$. Here $\chi_f$ is a \textit{signed} final spin with magnitude $|\bm{\chi}_f|$, and the same sign of $\bm{\chi}_f \cdot \bm{L}_f$.
This prescription of the postmerger extension of the Euler angles in the \emph{retrograde case} ($\bm{\chi}_f \cdot \bm{L}_f < 0$) is significantly less tested than in the prograde case due to the limited availability of NR simulations covering the relevant region of parameter space -- most notably high mass-ratio binaries -- which also includes systems with transitional precession~\cite{Apostolatos:1994mx}.

As mentioned earlier, the QNM frequencies obtained from BH perturbation theory are formally valid in the $\bm{J}_{\rm f}$-frame.
Following recent insights from NR~\cite{Hamilton:2023znn}, \seobfivephm{} computes the coprecessing frame QNM frequencies from the ones in the $\bm{J}_{\rm f}$-frame as [see Ref.~\cite{Ramos-Buades:2023ehm}, Eq.~(22) and Ref.~\cite{Hamilton:2023znn}, Eq.~(35)],
\begin{align}
  \omega^{\rm QNM,\,CP}_{\ell m 0} = \omega^{\rm QNM,\,J}_{\ell m 0}-m \left(1- |\cos \beta(t_{\rm match})|\right) \omega_{\rm prec}.
  \nonumber \\
\label{eq:qnm_CP}
\end{align}

\subsection{Construction of the parametrized waveform model}
\label{subsec:tgr_parameters}

The \pseob{} model introduces fractional deviation parameters to:
\begin{enumerate}
    \item the frequency and damping time of the fundamental QNM frequencies, \label{itm:qnms}
    \item the instant at which the GW amplitude peaks, the instantaneous GW frequency at this time instant, and the value of the peak amplitude, \label{itm:merger}
    \item two calibration parameters in the EOB Hamiltonian. \label{itm:hamilton}
\end{enumerate}
Items~\ref{itm:qnms} and~\ref{itm:merger} can be used independently for each mode of the model in Eq.~\eqref{eq:modes_v5hm}.
This yields a total of $31$ free parameters on top of the GR parameters. Specifically: there are $7\times2$ parameters related to the QNMs; $7 \times 2$ parameters related to the instantaneous GW frequency and amplitude; $1$ parameter associated with the instant the GW amplitude peaks; and $2$ EOB calibration parameters.

We introduce non-GR deformations to the QNMs, following the same strategy used in Refs.~\cite{Brito:2018rfr,Ghosh:2021mrv,Maggio:2022hre}. We perform the substitutions
\begin{subequations}
\begin{align}
    f_{\lm 0}^{\rm J} &\to f_{\lm 0}^{\rm J} \, (1 + \delta f_{\lm 0}),
    \\
    \tau_{\lm 0}^{\rm J} &\to \tau_{\lm 0}^{\rm J} \, (1 + \delta \tau_{\lm 0}), \label{tau}
\end{align}
\end{subequations}
where we impose that $\delta \tau_{\ell m 0} > -1$ to ensure that the remnant BH is
stable; it rings downs, instead of ``ringing-up'' exponentially.
In the following, we will always refer to frequencies in the $\bm{J}_{\rm f}$-frame, and drop the superscript $\mathrm{J}$ to lighten the notation.

If the binary is spin-aligned, the foregoing discussion is sufficient to
describe the parametrizations of the merger-ringdown part of the model.
However, if the binary is spin-precessing, we have to care about the different reference
frames used to model such systems.

In \texttt{pSEOBNRv5PHM}, we can add QNM deviations to the coprecessing frame waveform modes~\eqref{RD}, and
to the rotation angles that relate the $\bm{J}_{\rm f}$-frame and coprecessing frame after merger~\eqref{eq:euler_RD}.
As mentioned earlier, the QNM frequencies obtained from BH perturbation theory are formally valid in the $\bm{J}_{\rm f}$-frame. Therefore, it is most natural to add parametrized deviations to the frequencies in this frame.
We see that by deforming the QNM frequencies in the $\bm{J}_{\rm f}$-frame, we also change the effective precession rate $\omega_{\rm prec}$~\eqref{eq:omega_prec}, which gives an additional source of deviation from GR for the QNM frequencies in the coprecessing frame~\eqref{eq:qnm_CP}. Moreover, $\omega_{\rm prec}$ also appears in the phenomenological equations for the Euler angles after $t_{\rm match}$ that relate the waveform modes at coprecessing and $\bm{J}_{\rm f}$-frames.

At the transition between prograde-spin and retrograde-spin configurations, a
small discontinuity present in the rotations in the GR limit of the model is
amplified when including nonzero QNM deviations: the same deviation in $\delta f_{220}$,
or in $\delta f_{210}$, changes $\omega_{\rm prec}$ in opposite directions depending on the sign of $\bm{\chi}_f \cdot \bm{L}_f$ (see Eq.~\eqref{eq:omega_prec}).
This highlights the need to revisit the prescription for retrograde-spin configurations as more NR simulations covering the relevant parameter space become available~\cite{Hamilton:2023qkv}.
Here, to safely avoid this problem, we introduce a boolean parameter \textbf{\texttt{omega\_prec\_deviation}} that propagates (if \texttt{True}) or not (if \texttt{False}) the QNM modifications to $\omega_{\rm prec}$.
For PE applications, we choose to include QNM deviations in the rotations only if all posterior samples from a corresponding GR run are in a prograde-spin configuration.
For posteriors that are entirely in a negative-spin configuration, no discontinuity arises; however, we still prefer not to include QNM deviations in the rotations since the prescription has not been extensively validated against NR simulations. We also note that such configurations, characterized by high mass ratios and negative spins, are uncommon among the observed events.

A notable difference with respect to the aligned-spin case comes from mixing of modes with the same $\ell$ and different $m$ due to the rotation~\eqref{eq:prec_rotations}: when spin precession is present, even adding deviations to the coprecessing frame modes only, leads to corrections in a single QNM propagating to different modes with the same $\ell$ in the inertial frame.
Another source of mode-mixing -- this time present already in the aligned-spin model -- is the spherical-spheroidal mode mixing in the $(3,2)$ and $(4,3)$ modes. In this case, corrections from the $(2,2)$ and $(3,3)$ modes propagate to the $(3,2)$ and $(4,3)$ modes in a consistent way.

\begin{figure*}[ht]
    \includegraphics[width=\linewidth]{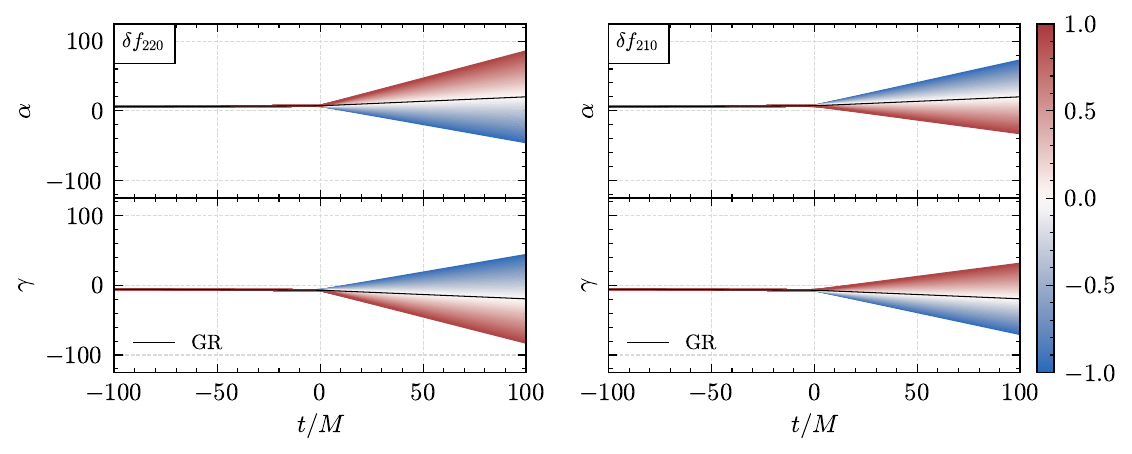}
	\caption{Time evolution near the merger of the $\alpha$ and $\gamma$ Euler angles for nonzero values of the QNM frequency deviations $\delta f_{220}$ (left panel) and $\delta f_{210}$ (right panel). The \gr predictions ($\delta f_{\ell m 0} = 0$) are shown by the black curves.}
	\label{fig:euler_angles}
\end{figure*}

Plunge-merger deviations are included similarly as in Ref.~\cite{Maggio:2022hre}.
We introduce fractional deviations to the NR-informed input values for the mode amplitudes and frequencies at $t = \tm$, i.e.,
\begin{subequations}
\label{eq:mod_h_and_omega}
\begin{align}
    |h_{\lm}^{\rm NR}| &\to |h_{\lm}^{\rm NR}| \, (1 + \delta A_{\lm})\,,
    \\
    \omega_{\lm}^{\rm NR} &\to \omega_{\lm}^{\rm NR} \, (1 + \delta \omega_{\lm})\,,
\end{align}
\end{subequations}
which are imposed via the NQC corrections. The factorized waveform modes and NQC corrections are directly used to compute coprecessing frame waveform during the inspiral-plunge phase~\eqref{eq:hlm_insp-plunge}. Consequently, the fractional deviations are applied to the amplitude and frequency of the modes in the coprecessing frame.

We also allow for changes to $\tm$ by modifying the time-shift parameter $\Delta t_{\rm NR}$, defined in Eq.~\eqref{eq:tISCO} as,
\begin{align}
    \Delta t_{\rm NR} \to \Delta t_{\rm NR} - \delta \Delta t \,.
\label{eq:mod_dt}
\end{align}
Unlike Ref.~\cite{Maggio:2022hre}, we introduce additive (rather than fractional) deviations to the peak time.
This choice is motivated by the different meaning of this parameter in the $\texttt{SEOBNRv4}$ and $\texttt{SEOBNRv5}$ models.
In $\texttt{SEOBNRv4}$ the time-shift parameter related to the peak of the $(2,2)$ mode to the peak of the orbital frequency, and had a constant sign across parameter space (i.e., the peak of the $(2,2)$ mode always occurs earlier than to peak of $\Omega$). In $\texttt{SEOBNRv5}$, however, the time-shift parameter relates the peak of the $(2,2)$ mode to the ISCO radius of the remnant Kerr BH, and can take both positive and negative values (i.e., the merger-ringdown attachment time is either before or after the ISCO), including zero, making fractional deviation ill defined in such cases.

Inspiral deviations modify EOB calibration parameters in the Hamiltonian, $a_6$ and $d_{\rm {SO}}$, as follows:
\begin{subequations}
    \label{eq:mod_h_and_omega}
    \begin{align}
        a_6 &\to a_6 + \delta a_6\,,
        \\
        d_{\rm{SO}} &\to d_{\rm{SO}} + \delta d_{\rm{SO}}\,.
    \end{align}
    \end{subequations}
Also in this case we apply parametrized corrections as additive deviations, since $a_6$ and $d_{\rm {SO}}$ can take on both positive and negative values across the parameter space.
Unlike previous corrections, which only affected the waveform modes without impacting the binary's orbital dynamics, these deviations modify the orbital dynamics in the coprecessing frame, and consistently propagate to all the waveform modes.

\begin{figure*}
    \includegraphics[width=\linewidth]{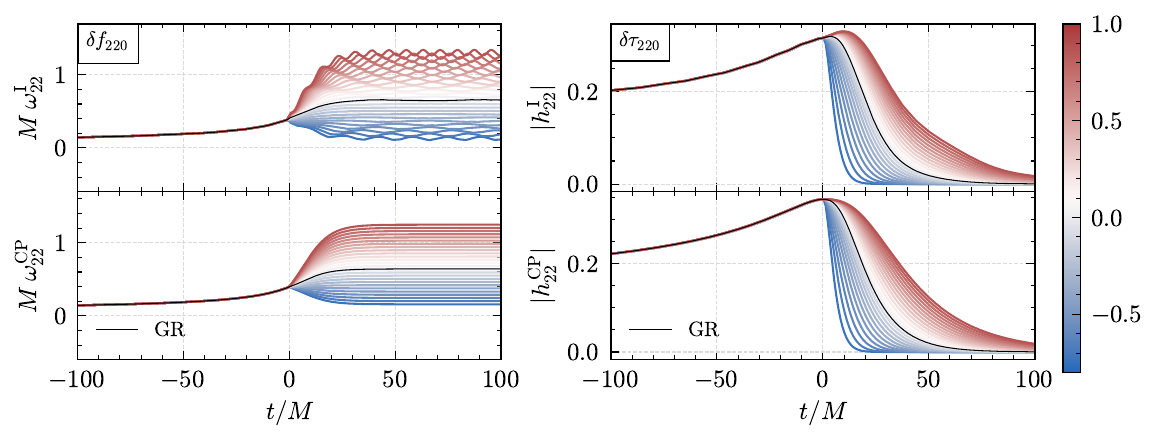}
    \includegraphics[width=\linewidth]{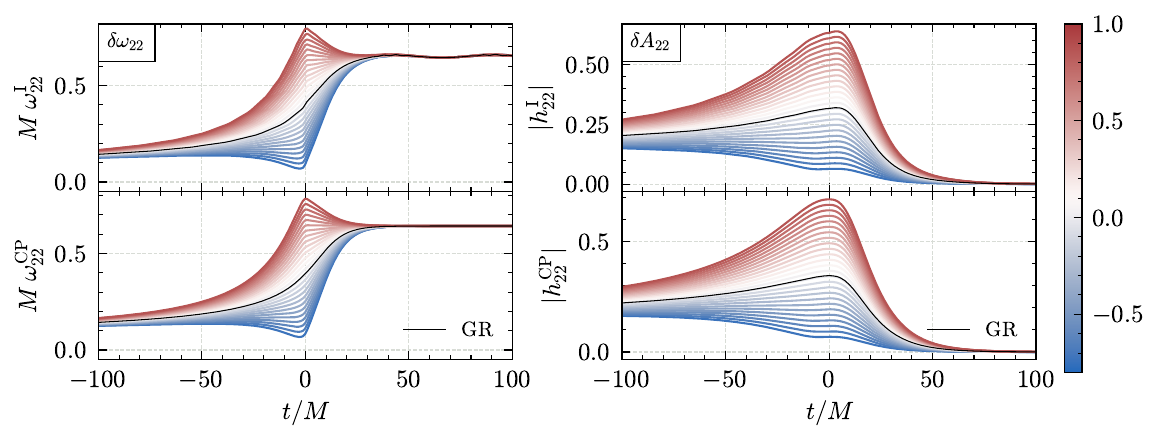}
    \caption{
        Impact of QNM and merger deviations on the waveform morphology. The top-left panel shows the effect of frequency deviations $\delta f_{220}$ on the instantaneous GW frequency in the inertial frame ($M \omega_{22}^{\rm I}$) and the coprecessing frame ($M \omega_{22}^{\rm CP}$). The top-right panel illustrates the effect of damping time deviations $\delta \tau_{220}$ on the waveform amplitude in both frames ($|h_{22}^{\rm I}|$ and $|h_{22}^{\rm CP}|$). The bottom-left and bottom-right panels explore the influence of merger frequency deviations $\delta \omega_{22}$ and merger amplitude deviations $\delta A_{22}$, respectively, on the GW frequency and amplitude in the inertial and coprecessing frames. Colored envelopes correspond to deviations sampled within the intervals $(\delta f_{220}, \delta \tau_{220}) \in [-0.8, 1.0]$ and $(\delta A_{22}, \delta \omega_{22}) \in [-0.8, 1.0]$, while black curves represent the GR prediction. The time $t=0$ is defined as the peak amplitude of the coprecessing $(2,2)$ mode. For precessing binaries, mode mixing is evident in the inertial frame, while the coprecessing frame exhibits a morphology consistent with aligned-spin binaries.
    }
	\label{fig:waveform_morphology}
\end{figure*}


\section{Morphology of the parametrized waveform}
\label{sec:morpho}

\begin{figure*}
    \includegraphics[width=\linewidth]{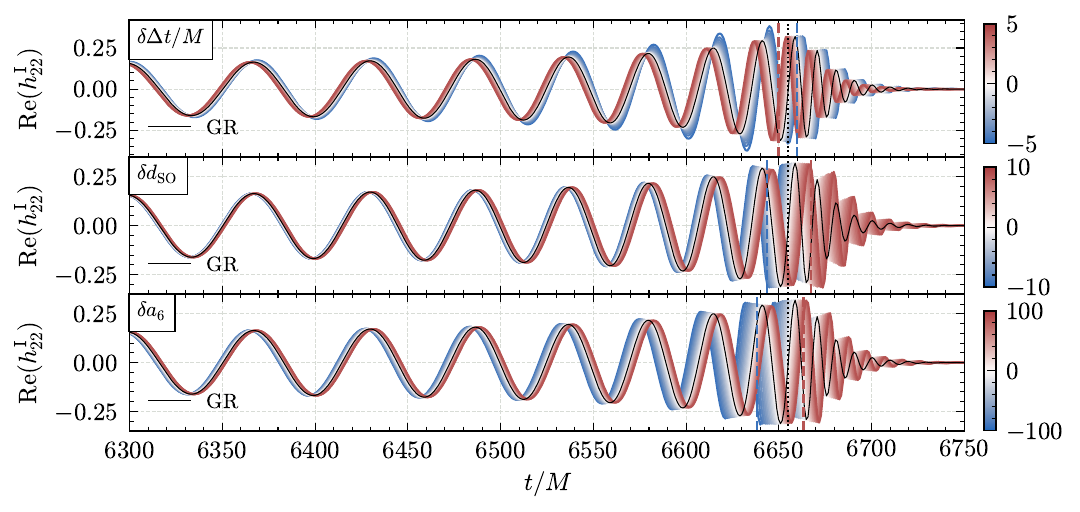}
    \caption{Impact of inspiral-plunge deviations on the waveform morphology. The top panel shows the effect of the time-shift correction $\delta \Delta t$, while the middle and bottom panel show, respectively, the effect of the $\delta d_{\rm{SO}}$ and $\delta a_6$ corrections to calibration parameters in the EOB Hamiltonian. Colored envelopes represent deviations sampled within the intervals $\delta \Delta t \in [-5, 5] M$, $\delta d_{\rm{SO}} \in [-10, 10]$, and $\delta a_6 \in [-100, 100]$, with the colored dashed lines denoting the merger-ringdown attachment time at the extremal values of the deviation parameters. Black curves correspond to the GR predictions, with the vertical dotted line marking the corresponding merger-ringdown attachment time.}
    \label{fig:waveform_morphology_insp}
\end{figure*}

Having introduced our waveform model, we now discuss how each of the parametrized deviations modifies the \gw signal in \gr, varying the parameters one at a time.
We follow analogous explorations performed for aligned-spin binaries in Ref.~\cite{Ghosh:2021mrv} for the ringdown stage, and in Ref.~\cite{Maggio:2022hre} for the plunge-merger stage.

As an illustrative example, we consider a spin-precessing quasicircular binary system with the following mass ratio and spins:
\begin{equation}
    q = 2.0, \quad \textrm{and} \quad
    \bm{\chi}_{1} = \bm{\chi}_{2} = [0.5, 0, 0.5]\,,
\label{eq:example_fixed_mass_spin}
\end{equation}
where the spins are defined at a dimensionless orbital frequency $M \Omega = 0.015$, corresponding to a frequency of $20~\rm{Hz}$ for a binary with total mass of approximately $50~M_{\odot}$.
We examine the impact of the deviations on both the rotation angles and the waveform in the coprecessing and inertial frames, focusing for clarity only on the $(2,2)$ mode.

Let us start with the QNM deviation parameters $\delta f_{\ell m 0}$ and $\delta \tau_{\ell m 0}$. Figure~\ref{fig:euler_angles} illustrates the impact of QNM frequency deviations $\delta f_{\ell m 0}$ on the postmerger extension of the Euler angles $\alpha$ (top rows) and $\gamma$ (bottom rows), as described by  Eq.~\eqref{eq:euler_RD}.
In the left panel, corrections are applied only to $f_{2 2 0}$, while in the right panel, corrections are applied only to $f_{2 1 0}$.

In all panels, the colored envelopes represent the range of deviations obtained by finely sampling the interval $\delta f_{\ell m 0} \in [-1.0, \,1.0]$ in increments of $0.04$, and the black curves correspond to the \gr angles ($\delta f_{\ell m 0}$ = 0) for the same binary parameters. The time $t=0$ corresponds to the peak of the coprecessing $(2,2)$ mode amplitude, which marks the attachment point of the merger-ringdown waveform.

The binary parameters in Eq.~\eqref{eq:example_fixed_mass_spin} describe a prograde-spin configuration ($\bm{\chi}_f \cdot \bm{L}_f > 0$). The morphology of the angles aligns with the expected behavior from Eqs.~\eqref{eq:omega_prec} and \eqref{eq:euler_RD}: a positive (negative) value of $\delta f_{220}$ increases (decreases) $\omega_{\rm prec}$, which subsequently increases (decreases) the slope of $\alpha$ and decreases (increases) the slope of $\gamma$ after the attachment time. Conversely, for $\delta f_{210}$, the opposite behavior is observed.

We now examine the impact of the QNM deviations on the waveform.
The top-left panel of Fig.~\ref{fig:waveform_morphology} focuses on the frequency deviations $\delta f_{2 2 0}$, while the top-right panel focuses on the damping time deviations $\delta \tau_{2 2 0}$.
For the frequency deviations, the top row shows the instantaneous GW frequency in the inertial frame ($M \omega_{22}^{\rm I}$), while the bottom row shows the frequency in the coprecessing frame ($M \omega_{22}^{\rm CP}$). For the damping time deviations, we instead examine the waveform amplitude in both inertial ($|h_{22}^{\rm I}|$) and coprecessing ($|h_{22}^{\rm CP}|$) frames.
The colored envelopes represent the range of deviations obtained by finely sampling the interval $(\delta f_{2 2 0},~\delta \tau_{220}) \in [-0.8, \,1.0]$ using 26 evenly spaced values, with the black curves corresponding to the \gr predictions. The time $t=0$ corresponds to the peak amplitude of the coprecessing $(2,2)$ mode.
For precessing binaries, modulations due to mode mixing between modes with the same $\ell$ in the coprecessing frame are evident in the inertial frame, particularly in the frequency. In the coprecessing frame, however, the morphology is analogous to that observed for aligned-spin binaries in previous studies~\cite{Ghosh:2021mrv}.
Overall, the parametrized model provides smooth deviations from GR, even in the inertial frame. As with aligned-spin binaries, QNM corrections affect the waveform only after the merger time, which corresponds to the peak of the coprecessing $(2,2)$ mode.

Turning to the merger deviations ($\delta A_{22}, \delta \omega_{2 2}$), the bottom-left panel of Fig.~\ref{fig:waveform_morphology} examines the merger frequency deviation $\delta \omega_{2 2}$, while the bottom-right panel focuses on the merger amplitude deviation $\delta A_{2 2}$. For merger frequency deviations, we plot the instantaneous GW frequency, while for merger amplitude deviations, we analyze the waveform amplitude. In both cases, the plots include data from both the inertial and coprecessing frames.
The deviation parameters are varied within the interval $(\delta A_{2 2},~\delta \omega_{22}) \in [-0.8, \,1.0]$ using 26 evenly spaced values.
In both inertial and coprecessing frames, the waveform phenomenology is consistent with that of aligned-spin binaries~\cite{Maggio:2022hre}. We note that for positive $\delta A_{2 2}$ the peak amplitude increases while maintaining a monotonic amplitude evolution. Conversely, for negative $\delta A_{2 2}$, the amplitude decreases, resulting in a local minimum bordered by two maxima located before and after the minimum.
Lastly, it is important to note that, although these parameters are referred to as ``merger parameters'', they also affect the late inspiral-plunge phase of the waveform (i.e., before the attachment time), with the NQC corrections ensuring a smooth transition across these phases.

Finally, Fig.~\ref{fig:waveform_morphology_insp} explores the impact of inspiral-plunge deviation parameters ($\delta \Delta t,\delta d_{\rm{SO}}, \delta a_6$) on the waveform morphology. In this figure, we present the real part of the $(2,2)$ mode in the inertial frame, $\mathrm{Re}(h_{22}^{\rm I})$.  These corrections primarily affect the time to merger and the late-inspiral phasing of the binary. To better highlight their effects, the waveforms are aligned in the early inspiral, rather than setting the peak of the coprecessing $(2,2)$ mode at $t=0$.
Colored envelopes represent deviations sampled within the intervals $\delta \Delta t \in [-5, 5] M$, $\delta d_{\rm{SO}} \in [-10, 10]$, and $\delta a_6 \in [-100, 100]$, using 26 evenly spaced values each. The colored dashed lines denote the merger-ringdown attachment time at the extremal values of the deviation parameters. Black curves correspond to the GR predictions, with the vertical dotted line marking the corresponding merger-ringdown attachment time.

The top panel shows the effect of the time-shift correction $\delta \Delta t$. Unlike Ref.~\cite{Maggio:2022hre}, we treat $\delta \Delta t$ as an additive, rather than fractional, deviation, allowing for a broader range of waveform morphologies.
The parameter $\delta \Delta t$ directly modifies the time at which the merger-ringdown waveform is attached~\eqref{eq:tISCO}, affecting the binary's time to merger. Negative (positive) values shift the merger to occur later (earlier), as dictated by the negative sign in Eq.~\eqref{eq:mod_dt}. While $\delta \Delta t$ does not alter the inspiral dynamics, it still impacts the inspiral-plunge amplitude and phasing of the waveform via NQC corrections applied at the shifted attachment time.
These corrections enforce the NR-calibrated amplitude and frequency at the attachment time, but their influence on the waveform before that point is nontrivial and depends sensitively on the time at which they are applied.
For example, we see that negative values of $\delta \Delta t$ may produce a maximum in amplitude before the attachment time.
The middle and bottom panels illustrate the effects of $\delta d_{\rm{SO}}$ and $\delta a_6$ respectively. Both parameters, as high-order PN corrections to the EOB Hamiltonian, exhibit qualitatively similar impacts on the phasing of the waveform during the late inspiral. Specifically, positive (negative) values delay (advance) the merger in time. The prefactor of $d_{\rm{SO}}$ depends on the effective spin parameter $\chi_{\rm eff}$ whereas $a_6$ is spin-independent. Consequently, the relative deviation with respect to GR caused by these corrections
varies across the parameter space. The ranges explored are indicative of possible PE priors, with the broader range of $\delta a_6$ compared to $\delta d_{\rm{SO}}$ reflecting results from NR calibration and corresponding uncertainties~\cite{Pompili:2023tna,Pompili:2024yec}.


\section{Parameter estimation on synthetic signals}
\label{sec:pe_injections}

In this section, we summarize the Bayesian inference formalism used for PE of \gw signals and synthetic-data studies.
Our PE studies focus on the QNM deviation parameters of the $(\ell,m,n)=(2,2,0)$ mode, as it is the primary test currently performed in \LVK{} analyses. In future work, we aim to extend this analysis to the higher modes in the ringdown, as well as the measurability of deviations in the inspiral and plunge-merger stages.

The goal of Bayesian PE is to infer the posterior distribution $p(\theta|d)$ for the parameters $\theta$ given the observed data $d$, using Bayes theorem
\begin{equation}
    P(\theta | d)=\frac{\mathcal{L}(d | \theta) \pi(\theta)}{Z},
\end{equation}
where $\mathcal{L}(d|\theta)$ is the likelihood of the data $d$ given the parameters $\theta$, $\pi(\theta)$ is the prior on $\theta$, and $Z \equiv \int \mathrm{d} \theta \mathcal{L}(d | \theta) \pi(\theta)$ is the evidence~\cite{Thrane:2018qnx}.
To determine whether a model $A$ is preferred over a model $B$, one can compute the Bayes factor, defined as the ratio of the evidence for the two different models $\mathcal{B}^A_B=Z_A/Z_B$.

We simulate and analyze signals using the {\tt Bilby} package~\cite{Ashton:2018jfp, Romero-Shaw:2020owr}, and the nested sampler \texttt{dynesty} \cite{Speagle:2019ivv} using the \texttt{acceptance-walk} stepping method. We adopt sampler settings consistent with the latest LVK analyses~\cite{LIGOScientific:2024elc} by using a number of accepted MCMC-chains $\mathrm{naccept}=60$, number of live points $\mathrm{nlive}=1000$, while keeping the remaining sampling parameters to their default values.
The \texttt{pSEOBNRv5PHM} waveforms are generated using \texttt{Bilby TGR}~\cite{ashton_2024_10940210} and \texttt{pySEOBNR}~\citep{Mihaylov:2023bkc}, interfaced through \texttt{gwsignal}.
When analyzing simulated signals, we consider a three-detector (LIGO Hanford, LIGO Livingston and Virgo) network configuration, and use the LIGO power spectral density (PSD) at O4 sensitivity~\cite{LIGO:NoiseCurves} and Virgo PSD at advanced Virgo sensitivity~\cite{KAGRA:2013rdx}. The noise curves are named \texttt{aLIGO\_O4\_high} and \texttt{AdV} in \texttt{Bilby}.

We use standard priors for the GR parameters following Refs.~\cite{Ramos-Buades:2023ehm, KAGRA:2021vkt}.
Specifically, we sample the masses using the chirp mass ($\mathcal{M}$) and inverse mass ratio ($1/q$), with priors uniform in component masses. The priors on the dimensionless spin vectors are uniform in magnitude $a_i \in [0, 0.99]$, and isotropically distributed in the unit sphere for the spin directions.
For the distance, we employ a prior uniform in the comoving-frame of the source, except in Sec.~\ref{subsec:NR_inj} where we use a prior uniform in distance to match the settings of the analysis in Ref.~\cite{Ramos-Buades:2023ehm} for the same simulated signal.
For the QNM deviation parameters, we use uniform priors in the ranges
\begin{equation}
    \delta f_{220} \in [-0.8, 2.0] \quad \textrm{and} \quad \delta \tau_{220} \in [-0.8, 2.0].
\end{equation}
In cases exhibiting railing, we extend the prior range to $[-0.8, 4.0]$.
The remaining priors are set according to Appendix C of Ref.~\cite{KAGRA:2021vkt}.

\subsection{Injection of a BBH signal in GR}
\label{subsec:NR_inj}

In this section, we assess the importance of including spin-precession effects when performing tests of GR, by analyzing a synthetic NR signal of a BBH in GR. The injected signal corresponds to the NR waveform {\tt SXS:BBH:0165} from the public SXS catalog~\cite{Boyle:2019kee}, with mass ratio $q=6$, detector-frame total mass $M=95\,M_\odot$ and BH's dimensionless spin vectors defined at 20~Hz of $\bm{\chi}_1 = [-0.06,0.78,-0.4]$ and $\bm{\chi}_2 = [0.08,-0.17,-0.23]$.
Notably, this BBH system exhibits strong spin precession, with a high mass ratio and a significant effective precessing-spin of $\chi_{\rm p} \sim 0.78$. It stands out as one of the most challenging systems to model in the public SXS catalog~\cite{Boyle:2019kee}.
We take the inclination with respect to the line of sight to be $\iota=\pi/2$ rad. The coalescence and polarization phases are $\phi=1.2$ rad and $\psi=0.7$ rad, respectively. The sky-position is defined by its right ascension of $0.33$ rad and declination of $-0.6$ rad at a geocentric time of $1249852257$ s. The luminosity distance to the source is chosen to be $1200$ Mpc, which produces a network SNR of $18.1$.

For this configuration, PE under the assumption of GR, using the \texttt{SEOBNRv5PHM} waveform model, yields posterior samples in both prograde-spin and retrograde-spin configurations. As discussed in Sec.~\ref{subsec:tgr_parameters}, at the transition between these configurations, corrections to $\delta f_{220}$ introduce a discontinuity in $\omega_{\rm prec}$. To address this, we exclude corrections to the $\omega_{\rm prec}$ term in such cases. Nevertheless, we analyze the signal under both approaches to evaluate the impact of this choice.

\begin{figure}
    \includegraphics[width=\columnwidth]{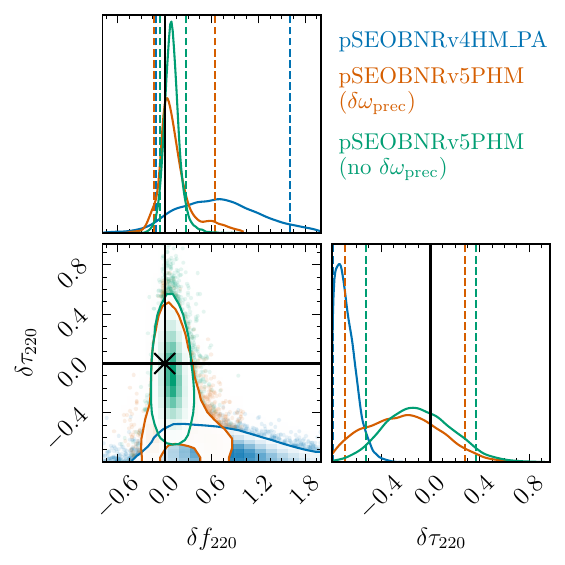}
	\caption{Posterior probability distributions for the fractional deviations in the frequency and damping time of the $(2,2,0)$ QNM ($\delta f_{2 2 0}$ and $\delta \tau_{2 2 0}$), for a synthetic signal of a highly precessing BBH NR waveform from the public SXS catalog \texttt{SXS:BBH:0165}.
    The PE is performed with the \texttt{pSEOBNRv4HM\_PA} model (blue) and with the \texttt{pSEOBNRv5PHM} model, including (orange) or excluding (green) the QNM corrections in the computation of the precession rate $\omega_{\rm prec}$~\eqref{eq:omega_prec}.
    The 2D contours mark the 90\% credible regions, while the dashed lines on the 1D marginalized distributions mark the 90\% credible levels. The black vertical and horizontal lines mark the GR predictions ($\delta f_{2 2 0} = \delta \tau_{2 2 0} = 0$).}
	\label{fig:corner_sxs_0165}
\end{figure}

Figure~\ref{fig:corner_sxs_0165} presents the posterior probability distributions for the fractional deviations in the frequency and damping time of the $(2,2,0)$ QNM ($\delta f_{2 2 0}$ and $\delta \tau_{2 2 0}$) for this synthetic signal.
We perform three parameter recoveries: with the \texttt{pSEOBNRv4HM\_PA} model~\cite{Maggio:2022hre, Mihaylov:2021bpf} (blue), and with the \texttt{pSEOBNRv5PHM} model, with (orange) and without (green) including QNM corrections in the computation of the precession rate $\omega_{\rm prec}$~\eqref{eq:omega_prec}. The 2D contours mark the 90\% credible regions, while the dashed lines on the 1D marginalized distributions mark the 90\% credible levels. The black vertical and horizontal lines mark the GR predictions ($\delta f_{2 2 0} = \delta \tau_{2 2 0} = 0$).

First, we observe that the \texttt{pSEOBNRv4HM\_PA} model, which does not include spin-precession effects, recovers a significant bias in the damping time away from the GR prediction. While some GR parameters (e.g., the masses) are still well-recovered, some parameters such as the binary's inclination are significantly biased.
As highlighted in previous studies~\cite{Maggio:2022hre, Chandramouli:2024vhw}, waveform systematics, due to the absence of spin-precession effects, can lead to false indications of deviations from GR.
On the other hand, at current detector sensitivity, the \texttt{pSEOBNRv5PHM} model demonstrates sufficient accuracy to recover GR predictions reliably, even for challenging binary configurations.
This is true both with and without including QNM corrections in $\omega_{\rm prec}$, although posterior distributions in the latter case appear more sensible, consistently with the avoidance of a discontinuous limit. This choice has no significant impact on other parameters, which are accurately recovered in both configurations of \texttt{pSEOBNRv5PHM}.

The natural log Bayes factor between \texttt{pSEOBNRv5PHM} without and with QNM corrections in $\omega_{\rm prec}$ is $\ln \mathcal{B} \simeq 0.2$, which is comparable with the estimated error in the Bayes factor, indicating no significant preference over either model. On the other hand, the natural log Bayes factor between \texttt{pSEOBNRv5PHM} without QNM corrections in $\omega_{\rm prec}$ and \texttt{pSEOBNRv4HM} is $\ln \mathcal{B} \simeq 5.7$ indicating a significant preference for the \texttt{pSEOBNRv5PHM} model. Finally, the natural log Bayes factor between \texttt{pSEOBNRv5PHM} without QNM corrections in $\omega_{\rm prec}$ and \texttt{SEOBNRv5PHM} in GR is $\ln \mathcal{B} \simeq -1.5$, indicating overall preference for the GR hypothesis.

In spin-precessing binaries, the relative amplitudes of different modes can be reordered compared to the nonprecessing scenario. In certain regions of the parameter space, the $(2,2,0)$ mode may become subdominant, while modes with $\ell = m \neq 2$ can exhibit large amplitudes~\cite{Hughes:2019zmt, Lim:2022veo, Siegel:2023lxl, Zhu:2023fnf}. This observation motivates the investigation of whether the $(2,1,0)$ mode can also be effectively constrained. For the \texttt{SXS:BBH:0165} NR injection, we perform a recovery that allows for deviations in both the $(2,2,0)$ and $(2,1,0)$ QNMs. The results indicate that the posterior distribution for the $(2,1,0)$ mode is only marginally informative: the damping time excludes large positive deviations, while the frequency remains compatible with the entire prior range, suggesting that, for this specific NR simulation, the $(2,1,0)$ mode remains subdominant relative to the $(2,2,0)$ mode.

While more detailed studies on model accuracy for future detectors remain to be conducted, we anticipate that improvements to the baseline GR model will be necessary for robust applications to LISA and next-generation ground-based detectors. This expectation aligns with similar findings for aligned-spin systems, especially for signals with SNRs reaching 100 or higher~\cite{Toubiana:2023cwr}.

\subsection{Injection of a beyond-GR signal}

In this section, we study whether our model can identify a signal that does not originate from a BBH in GR.

\begin{figure}
    \includegraphics[width=\columnwidth]{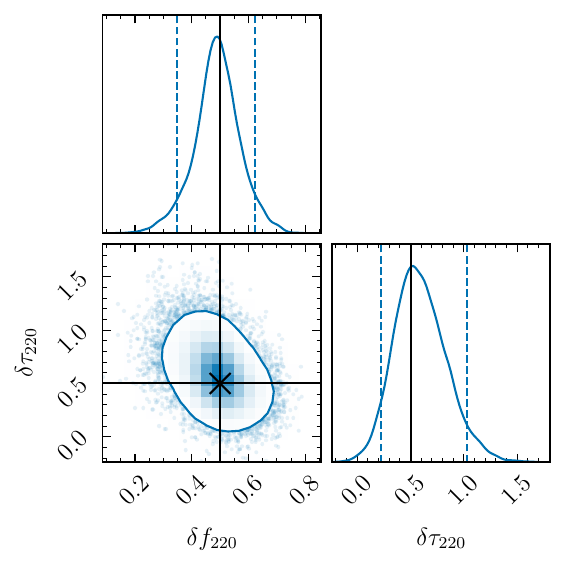}
	\caption{Posterior probability distributions for the fractional deviations in the frequency and damping time of the $(2,2,0)$ QNM ($\delta f_{2 2 0}$ and $\delta \tau_{2 2 0}$), for a synthetic signal modeled using \texttt{pSEOBNRv5PHM} with nonzero values for the QNM deviation parameters, and recovered with the same model. The 2D contours mark the 90\% credible regions, while the dashed lines on the 1D marginalized distributions mark the 90\% credible intervals. The black vertical and horizontal lines mark the injected values ($\delta f_{2 2 0} = \delta \tau_{2 2 0} = 0.5$). }
	\label{fig:corner_model_inj}
\end{figure}

In our first example, we simulate a signal using the \texttt{pSEOBNRv5PHM} model with nonzero QNM deviation parameters, $\delta f_{2 2 0} = \delta \tau_{2 2 0} = 0.5$. The simulated binary is a mass-asymmetric BBH with moderate spin-precession, with parameters $q=4,~(1+z)M = 75~M_{\odot},~\chi_{\mathrm{eff}}\simeq 0.15,~\chi_{\mathrm{p}}\simeq 0.6$, and a network SNR of $19.1$.
Figure~\ref{fig:corner_model_inj} shows the posterior probability distributions for the fractional deviations in the frequency and damping time of the $(2,2,0)$ QNM, recovered with \texttt{pSEOBNRv5PHM}. The 2D contours mark the 90\% credible regions, while the dashed lines on the 1D marginalized distributions mark the 90\% credible levels. The black vertical and horizontal lines mark the injected values ($\delta f_{2 2 0} = \delta \tau_{2 2 0} = 0.5$), which are accurately recovered.
The GR parameters are also well estimated. In contrast, recovering the same signal with the GR model \texttt{SEOBNRv5PHM} leads to biased intrinsic parameter estimates. In this case, the analysis incorrectly favors equal masses and a negative effective spin $\chi_{\mathrm{eff}}\simeq -0.4$, with the injected values falling outside the 90\% credible intervals. The natural log Bayes factor between \texttt{pSEOBNRv5PHM} and \texttt{SEOBNRv5PHM} is $\ln \mathcal{B} \simeq 15.7$, indicating strong evidence for a deviation from GR.

\subsection{Injection of a binary boson-star signal}

As a second example, we consider a synthetic signal from a publicly available NR simulation of a scalar-field solitonic boson star (BS)~\cite{Liebling:2012fv} merger, produced with the \texttt{GRChombo} code~\cite{Clough:2015sqa, Andrade:2021rbd} and described in Ref.~\cite{Evstafyeva:2024qvp}, where several high-precision, IMR waveforms spanning approximately 20 orbits were presented for equal-mass, quasicircular, nonspinning BS binaries of different compactness.

Boson stars 
are modeled by a complex scalar field, which can be decomposed into amplitude $A$ and frequency $\omega$, as $\varphi(t, r)=A(r) e^{\mathrm{i}(\epsilon \omega t+\delta \phi)}$.
Following Ref.~\cite{Evstafyeva:2024qvp} we introduce the parameter $\epsilon=\pm1$ determining the rotation of the scalar field in the complex plane, and a phase offset $\delta \phi$. The primary BS always has $\epsilon=1$, $\delta \phi=0$,  and we consider the configuration with secondary parameters $\epsilon=1$ and $\delta \phi=\pi$ (\textit{antiphase}). The simulation we consider has central amplitude $A(0) = 0.17$, which gives rise to highly compact BSs with compactness $\mathcal{C}=0.2$ and dimensionless tidal deformability $\Lambda \sim10$. The coalescence results in the formation of a BH with final spin $\chi_f\simeq 0.7$, similar to the remnant of a nonspinning equal-mass BBH merger.

The injection setup matches the one for NR injections of BBHs, but we include only the $\ell=2$ modes in both injection and recovery, since they are the only ones contained in the NR data. We take the same total mass and extrinsic parameters as for the \texttt{SXS:BBH:0165} NR injection, except for the inclination angle which we take $\iota=\pi/3$ rad to give an optimal SNR of $31.5$.

\begin{figure*}
    \includegraphics[height=8.5cm, keepaspectratio]{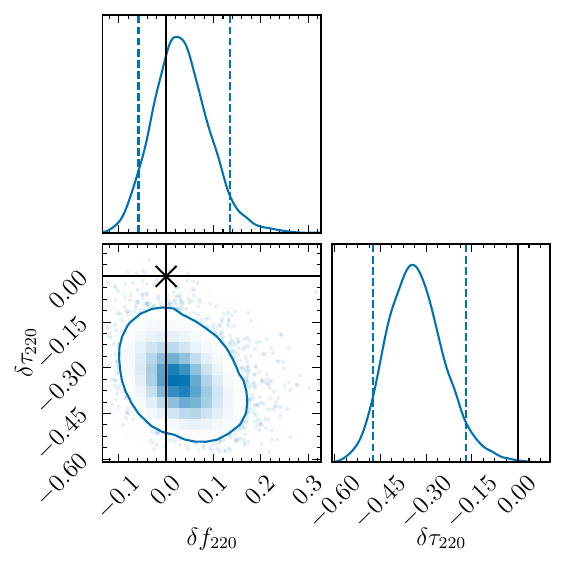}
    \includegraphics[height=8.5cm, keepaspectratio]{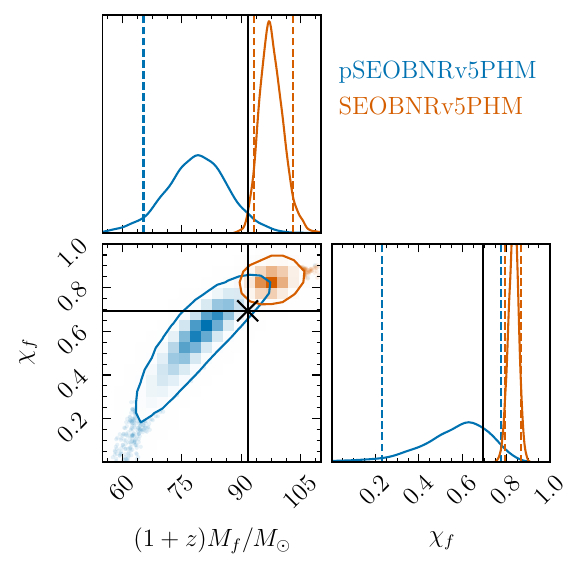}
	\caption{
        \emph{Left panel}: posterior probability distributions for the fractional deviations in the frequency and damping time of the $(2,2,0)$ QNM ($\delta f_{2 2 0}$ and $\delta \tau_{2 2 0}$), for a synthetic signal of a binary BS merger recovered with \texttt{pSEOBNRv5PHM}. The 2D contours mark the 90\% credible regions, while the dashed lines on the 1D marginalized distributions mark the 90\% credible intervals. The black vertical and horizontal lines mark the predictions for a BBH in GR ($\delta f_{2 2 0} = \delta \tau_{2 2 0} = 0$).
        \emph{Right panel}: posterior probability distributions for the detector-frame mass and spin of the remnant BH, estimated with the \texttt{pSEOBNRv5PHM} and \texttt{SEOBNRv5PHM} models. For \texttt{pSEOBNRv5PHM}, the remnant properties are estimated from the complex QNM frequencies by inverting the fitting formula provided in Ref.~\cite{Berti:2005ys}. The black vertical and horizontal lines mark the true values.
        }
	\label{fig:corner_boson_star}
\end{figure*}

The left panel of Fig.~\ref{fig:corner_boson_star} shows the posterior probability distributions for the fractional deviations in the frequency and damping time of the $(2,2,0)$ QNM ($\delta f_{2 2 0}$, $\delta \tau_{2 2 0}$) for the synthetic binary BS signal recovered with \texttt{pSEOBNRv5PHM}.
The prediction for a BBH in GR ($\delta f_{2 2 0} = \delta \tau_{2 2 0} = 0$) is excluded at the 90\% credible level, demonstrating that \texttt{pSEOBNRv5PHM} successfully identifies a deviation from the signal of a BBH in GR. Specifically, while $\delta f_{2 2 0}$ remains consistent with zero, the posterior for $\delta \tau_{2 2 0}$ shows a preference for negative values, indicating a more strongly damped ringdown signal
compared to what would be predicted from the component masses and spins of a BBH merger remnant that matches the inspiral signal. 

When the BS signal is analyzed under the assumption of a BBH in GR, the reconstruction of the ringdown can be biased, because the inferred component masses and spins are themselves biased, and the final mass and spin are computed assuming a remnant from a BBH merger.
For the simulation considered, the bias in the component masses and spins is the larger source of error, as the mass and spin of the remnant BH are nearly identical to those of a BBH merger remnant.
The biases in the final mass and spin lead to incorrect estimates of the ringdown frequency and damping time. By allowing for corrections to the QNM frequencies, the \texttt{pSEOBNR} model is partially able to mitigate these biases and provide a better fit to the BS simulation.
This is confirmed by the right panel of Fig.~\ref{fig:corner_boson_star}, which shows the detector-frame mass and spin of the remnant BH, estimated with the \texttt{pSEOBNRv5PHM} and \texttt{SEOBNRv5PHM} models. For \texttt{pSEOBNRv5PHM}, the remnant properties are estimated from the complex QNM frequencies by inverting the fitting formula provided in Ref.~\cite{Berti:2005ys}.
The \texttt{SEOBNRv5PHM} estimate is not compatible with the true values, indicated by the black vertical and horizontal lines. The inclusion of a damping time deviation in the \texttt{pSEOBNRv5PHM} model shifts the posterior toward the true values, particularly for the final spin, which are now recovered within the 90\% credible region.

Note that the \texttt{pSEOBNR} model can infer that the signal does not originate from a BBH coalescence, even if the remnant is a BH in GR. In contrast, ringdown analyses limited to the postmerger stage might not reveal such discrepancies unless the ringdown results are compared to those from a complete IMR analysis based on the BBH assumption. In that scenario, the deviation could manifest either as a discrepancy between the recovered frequency and damping time and those predicted by the IMR analysis (when using an agnostic damped-sinusoid model) or as a mismatch between the IMR-inferred remnant mass and spin and those obtained from the ringdown stage (when employing a ringdown model that assumes a Kerr remnant).

The Bayes factor against the null hypothesis $\delta f_{2 2 0} = \delta \tau_{2 2 0} = 0$ can be estimated using the Savage-Dickey density ratio~\cite{10.1214/aoms/1177693507} without requiring an additional \texttt{SEOBNRv5PHM} run:
\begin{equation} \label{eq:savage}
    \mathcal{B} = \frac{\pi(\delta f_{2 2 0} = \delta \tau_{2 2 0} = 0)}{P(\delta f_{2 2 0} = \delta \tau_{2 2 0} = 0 | d)}.
\end{equation}
For uniform priors in the range $[-0.8, 2.0]$, for both $\delta f_{2 2 0}$ and  $\delta \tau_{2 2 0}$, Eq.~\eqref{eq:savage} yields a natural log Bayes factor of $\ln \mathcal{B} = 0.86$. It is important to note that the exact value of the Bayes factor depends on the choice of prior for the deviation parameters, which is somewhat arbitrary. For instance, using priors uniform in $[-0.8, 1.0]$ instead, the natural log Bayes factor becomes $\ln \mathcal{B} = 1.74$. These results indicate a slight preference for a deviation from the null hypothesis, despite the inclusion of additional parameters.

We also observe biases in the GR parameters consistent with results reported in Ref.~\cite{Evstafyeva:2024qvp}. Specifically, the posteriors show significant support away from equal masses, where the true value $q=1$ is excluded at the 90\% credible level, leading to an overestimation of the primary mass. The spin magnitudes exhibit substantial support for high values ($a_1, a_2 > 0.5$), with partial alignment to the orbital angular momentum ($\chi_{\rm eff} \simeq 0.5$). These biases in the mass ratio and effective spin allow the model to better reproduce the shallow chirp in the BS signal during the inspiral phase. Additionally, the luminosity distance is biased toward larger values.

\begin{table*}[t]
    \centering
    \begin{tabular}{lcccccc}
\toprule
Event & $\delta f_{220}$ & $\delta \tau_{220}$ & $\fngr{220}$ (Hz) & $\taungr{220}$ (ms) & $(1+z)M_f/M_{\odot}$ & $\chi_f$ \\[0.095cm]
\midrule

GW150914 &
$0.02^{+0.09}_{-0.07}$ &
$0.12^{+0.33}_{-0.27}$ &
$240.5^{+25.1}_{-25.6}$ &
$4.48^{+1.29}_{-1.06}$ &
$72.9^{+12.3}_{-12.8}$ &
$0.72^{+0.13}_{-0.28}$
\\[0.095cm]

GW170104 &
$-0.02^{+0.14}_{-0.13}$ &
$0.43^{+1.01}_{-0.66}$ &
$296.6^{+58.9}_{-54.5}$ &
$5.04^{+3.76}_{-2.37}$ &
$69.9^{+16.2}_{-20.6}$ &
$0.87^{+0.09}_{-0.42}$
\\[0.095cm]

GW190519\_153544 &
$-0.14^{+0.20}_{-0.13}$ &
$0.17^{+0.48}_{-0.35}$ &
$120.4^{+17.2}_{-18.4}$ &
$8.36^{+3.96}_{-2.65}$ &
$140.7^{+34.8}_{-31.2}$ &
$0.67^{+0.20}_{-0.45}$
\\[0.095cm]

GW190521\_074359 &
$0.06^{+0.17}_{-0.10}$ &
$-0.03^{+0.35}_{-0.26}$ &
$198.5^{+30.6}_{-28.9}$ &
$5.41^{+1.82}_{-1.41}$ &
$88.4^{+14.9}_{-16.6}$ &
$0.72^{+0.16}_{-0.37}$
\\[0.095cm]

GW190630\_185205 &
$-0.06^{+0.12}_{-0.16}$ &
$0.00^{+0.56}_{-0.45}$ &
$247.6^{+34.9}_{-44.2}$ &
$3.96^{+2.32}_{-1.76}$ &
$69.4^{+16.5}_{-17.9}$ &
$0.70^{+0.20}_{-0.54}$
\\[0.095cm]

GW190828\_063405 &
$0.11^{+0.11}_{-0.12}$ &
$0.18^{+0.52}_{-0.45}$ &
$226.7^{+40.1}_{-41.3}$ &
$6.18^{+2.67}_{-2.40}$ &
$88.4^{+15.6}_{-20.1}$ &
$0.85^{+0.09}_{-0.37}$
\\[0.095cm]

GW190910\_112807 &
$0.01^{+0.11}_{-0.09}$ &
$0.60^{+0.63}_{-0.47}$ &
$175.0^{+23.7}_{-20.1}$ &
$9.45^{+3.48}_{-2.67}$ &
$122.6^{+18.1}_{-18.6}$ &
$0.90^{+0.05}_{-0.12}$
\\[0.095cm]

GW191109\_010717 &
$1.31^{+0.65}_{-1.26}$ &
$-0.06^{+0.82}_{-0.53}$ &
$162.7^{+97.5}_{-80.3}$ &
$13.67^{+16.67}_{-10.46}$ &
$147.0^{+118.6}_{-72.3}$ &
$0.94^{+0.04}_{-0.39}$
\\[0.095cm]

GW200129\_065458 &
$-0.01^{+0.06}_{-0.07}$ &
$0.18^{+0.42}_{-0.29}$ &
$259.4^{+30.0}_{-23.0}$ &
$5.30^{+1.97}_{-1.35}$ &
$76.5^{+11.0}_{-10.9}$ &
$0.85^{+0.08}_{-0.19}$
\\[0.095cm]

GW200208\_130117 &
$0.25^{+1.65}_{-0.35}$ &
$-0.07^{+1.10}_{-0.43}$ &
$215.0^{+131.8}_{-56.9}$ &
$5.06^{+10.90}_{-2.33}$ &
$80.9^{+32.8}_{-25.6}$ &
$0.76^{+0.23}_{-0.56}$
\\[0.095cm]

GW200224\_222234 &
$0.01^{+0.15}_{-0.11}$ &
$0.22^{+0.46}_{-0.33}$ &
$206.2^{+25.4}_{-18.4}$ &
$7.07^{+2.76}_{-1.94}$ &
$98.9^{+13.0}_{-15.2}$ &
$0.87^{+0.08}_{-0.17}$
\\[0.095cm]

GW200311\_115853 &
$0.01^{+0.15}_{-0.07}$ &
$0.29^{+1.57}_{-0.54}$ &
$256.2^{+32.3}_{-24.3}$ &
$5.99^{+6.78}_{-2.55}$ &
$81.6^{+21.9}_{-21.4}$ &
$0.88^{+0.09}_{-0.35}$
\\[0.095cm]

\bottomrule
\end{tabular}
    \caption{The median and symmetric 90\% credible intervals of the one-dimensional marginalized posteriors of the fractional deviations in the frequency and damping time of the $(2, 2,0)$ QNM, $(\delta f_{220},\delta \tau_{220})$, and of the remnant properties. The third and fourth columns list the frequency and damping time of the $(2,2,0)$ QNM, as measured using the \texttt{pSEOBNRv5PHM} model. The last two columns report the mass and spin of the remnant object, estimated from the complex QNM frequencies by inverting the fitting formula provided in Ref.~\cite{Berti:2005ys}.
    }
    \label{tab:qnm_o1o2o3_results}
\end{table*}

\section{Parameter estimation on real data}
\label{sec:pe_real_data}

\begin{figure*}[t]
    \includegraphics[width=0.5\textwidth]{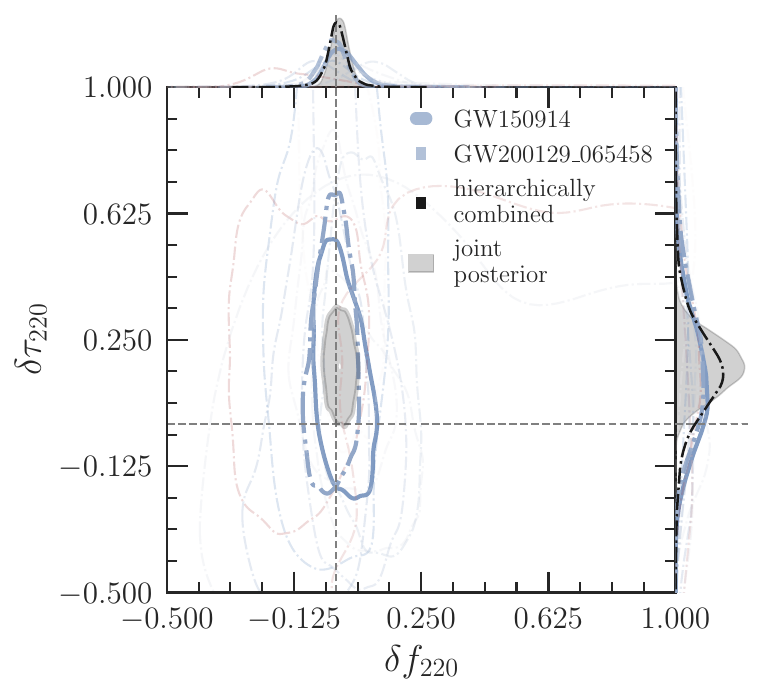}
    \includegraphics[width=0.3\textwidth]{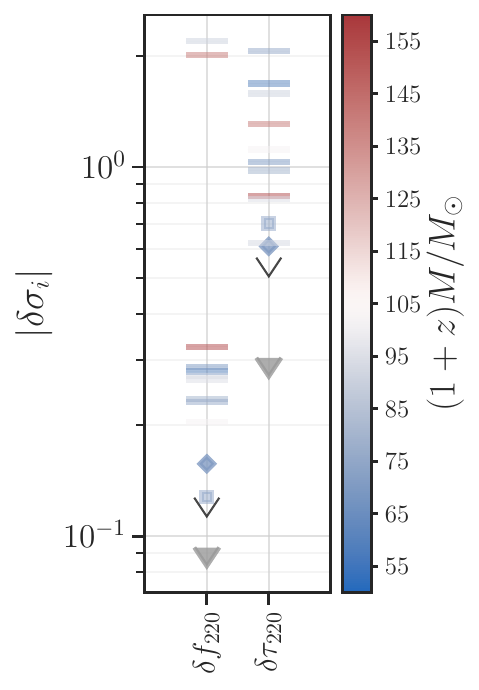}
    \caption{
        \emph{Left panel}: the 90\% credible regions of the posterior probability distribution of the fractional deviations in the frequency and damping time of the $(2, 2, 0)$ QNM, $\delta f_{2 2 0}$ and $\delta \tau_{2 2 0}$, and their corresponding one-dimensional marginalized posterior distributions, for events from GWTC-3 passing a SNR threshold of $8$ in both the inspiral and postinspiral signal.
        Posteriors for GW150914 and GW200129\_065458 are separately shown.
        The filled gray contours denote the 90\% credible regions on the joint constraints for $(\df{220},\dtau{220})$ obtained by multiplying the individual event posteriors (given a flat prior), while the hierarchical method of combination yields the black dot dashed curves only shown in the 1D marginalized posteriors. The dashed gray lines mark the GR prediction $(\delta f_{220},\delta \tau_{220})=(0,0)$.
        \emph{Right panel}: 90\% credible interval on the one-dimensional marginalized posteriors on $\delta \sigma_i=(\df{220},\dtau{220})$, colored by the median detector-frame total mass $(1 + z)M$, inferred assuming GR. Filled gray (unfilled black) triangles mark the constraints obtained when all the events are combined by multiplying likelihoods (hierarchically). The bounds from GW200129\_065458 (square) and GW150914 (diamond) are indicated by the separate markers.
    }
    \label{fig:o1o2o3_events}
\end{figure*}

In this section, we apply our model to real data by reanalyzing 12 events from GWTC-3, which were originally analyzed using the \texttt{pSEOBNRv4HM} model~\cite{Ghosh:2021mrv} in Ref.~\cite{LIGOScientific:2021sio}.

For the \texttt{pSEOBNR} test, a degeneracy exists between the fundamental ringdown frequency deviation parameter and the remnant mass~\cite{Ghosh:2021mrv}, mostly for low-SNR events with negligible higher modes and for which only the postmerger is detectable. To mitigate this, a selection criterion is applied, requiring that both the inspiral and postinspiral regimes achieve \snr{} $> 8$, a criterion met by 12 binary systems from GWTC-3 that also satisfy the other selection criteria for tests of GR (detection in multiple interferometers and false-alarm rates $< 10^{-3}~\mathrm{yr}^{-1}$). The final list of events used for this analysis is provided in Table~\ref{tab:qnm_o1o2o3_results}. We employ strain data from the Gravitational Wave Open Source Catalog (GWOSC)~\cite{LIGOScientific:2019lzm} and the released PSD and calibration envelopes.

As in previous studies~\cite{LIGOScientific:2019fpa, LIGOScientific:2020tif, LIGOScientific:2021sio}, we present results combining information from multiple events, as this allows to place stronger constraints on the deviation parameters.
Assuming that the fractional deviations $(\df{220},\dtau{220})$ are constant across events, joint constraints can be obtained by multiplying the individual-event posteriors (given a flat prior on the deviation parameters)~\cite{Zimmerman:2019wzo, LIGOScientific:2019fpa}.
However, in most non-GR theories, the deviations parameters $(\df{220},\dtau{220})$ are expected to vary depending on the source's properties.
Relaxing the assumption of constant deviations across all events requires a hierarchical inference framework, as originally proposed in Refs.~\cite{Zimmerman:2019wzo,Isi:2019asy}. This technique assumes that the non-GR parameters $(\df{220},\dtau{220})$ are drawn from a common underlying distribution, whose properties are inferred from the population of events. Following Refs.~\cite{Zimmerman:2019wzo,Isi:2019asy,LIGOScientific:2020tif,LIGOScientific:2021sio} we model the population distribution with a Gaussian  $\mathcal{N}(\mu,\sigma)$ of unknown mean $\mu$ and standard deviation $\sigma$ (the \textit{hyperparameters}).
The goal is then to infer a posterior distribution $P(\mu, \sigma |  \{d_j\}) $ for $\mu$ and $\sigma$ from a joint analysis of a set of events $\{d_j\}$, $j=1, ..., N$. If GR is correct, this posterior should be consistent with $\mu=0$ and $\sigma=0$.
To perform this analysis, we use the \texttt{stan}-based code~\cite{stan} developed in Ref.~\cite{Isi:2019asy} and employed in Refs.~\cite{Ghosh:2021mrv, LIGOScientific:2020tif, LIGOScientific:2021sio}. This code allows us to infer $P(\mu, \sigma |  \{d_j\}) $ across a set of events.
From a posterior on the hyperparameters one can also infer population distributions for the original deviation parameters $(\df{220},\dtau{220})$, by marginalizing over $\mu$ and $\sigma$.
However, as originally developed in Ref.~\cite{Isi:2019asy}, this implementation is defined only for 1D posteriors. Therefore, we restrict our presentation below to 1D hierarchical analyses (see Ref.~\cite{Zhong:2024pwb} for a recent extension to multidimensional cases).
Incorporating information about the underlying astrophysical population, such as simultaneously inferring the astrophysical population in the hierarchical analysis, would be important to mitigate the impact of unphysical prior assumptions on astrophysical parameters~\cite{Payne:2023kwj}, which can impact non-GR deviations due to parameter correlations.
However, we leave such an extension for future work, and use the same setup as Refs.~\cite{Ghosh:2021mrv, LIGOScientific:2021sio}.

The events GW191109\_010717 and GW200208\_130117 were not included in the computation of the combined bounds (hierarchical or joint posterior) in Ref.~\cite{LIGOScientific:2021sio}, as the posteriors on $\delta f_{220}$ show multimodalities likely due to possibility of noise systematics not accounted for. We performed single-event analyses also for these events, finding consistent results with Ref.~\cite{LIGOScientific:2021sio}. Therefore, we also do not include them in the combined results.

The results of the analysis are summarized in Fig.~\ref{fig:o1o2o3_events}, which is based on Fig.~14 of Ref.~\cite{LIGOScientific:2021sio}. The left panel of Fig.~\ref{fig:o1o2o3_events} shows the 2D posteriors (along with the marginalized 1D posteriors) of the frequency and damping time deviations for all the events listed in Table~\ref{tab:qnm_o1o2o3_results}. The contours are colored by the median detector-frame total mass $(1 + z)M$ of the corresponding binary. We specifically highlight the posteriors from two events, GW150914 and GW200129\_065458, which are among the loudest detected so far and provide strongest single-event bounds. The combined constraints are reported both by multiplying individual posteriors and by hierarchically combining events.
In the right panel of Fig.~\ref{fig:o1o2o3_events} we also provide a summary of the 90\% credible intervals on the 1D marginalized posteriors, color coded by the median detector-frame mass of the binary.

The results for GW150914 are broadly consistent with those reported in Ref.~\cite{LIGOScientific:2021sio}, with a Bayes factor between \texttt{pSEOBNRv5PHM} and \texttt{pSEOBNRv4HM} $\ln \mathcal{B} \simeq 0.6$.~\footnote{The original results with the \texttt{pSEOBNRv4HM} model from Ref.~\cite{LIGOScientific:2021sio} were produced using the \texttt{LALInference} code~\cite{Veitch:2014wba} and do not report an estimate of the Bayesian evidence. For GW200129\_065458 and GW150914 the Bayes factors are estimated from re-runs using \texttt{Bilby} and \texttt{dynesty}, employing identical settings and priors as the \texttt{pSEOBNR5PHM} model (except for the spins, which are aligned), but using the \texttt{pSEOBNRv4HM\_PA} model. }
This is expected, as GW150914 is consistent with originating from a nonspinning binary, making the impact of waveform systematics subdominant~\cite{LIGOScientific:2016ebw}.
In contrast, the posterior for GW200129\_065458 differs more noticeably from the one in Ref.~\cite{LIGOScientific:2021sio}. This event exhibits evidence of spin precession under the assumption of a binary in a quasicircular orbit~\cite{hannam:2021pit}. However, uncertainties in glitch subtraction could affect the evidence for spin precession~\cite{payne:2022spz, Macas:2023wiw}, and an alternative interpretation as an aligned-spin eccentric binary has been proposed~\cite{Gupte:2024jfe}.

\begin{figure*}[t]
    \centering
    \includegraphics[height=5.4cm, keepaspectratio]{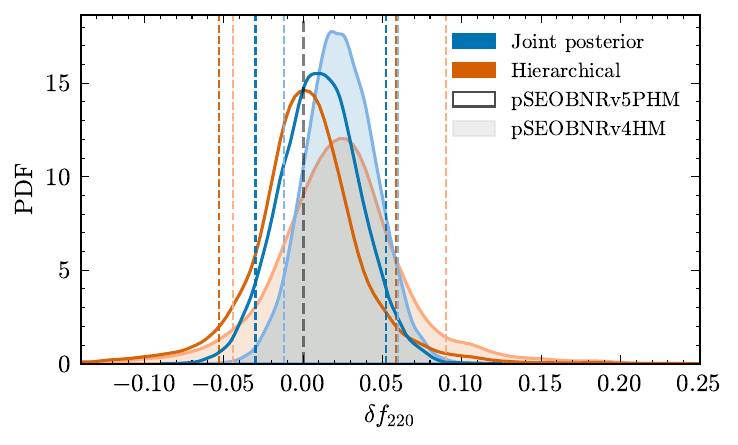}
    \includegraphics[height=5.4cm, keepaspectratio]{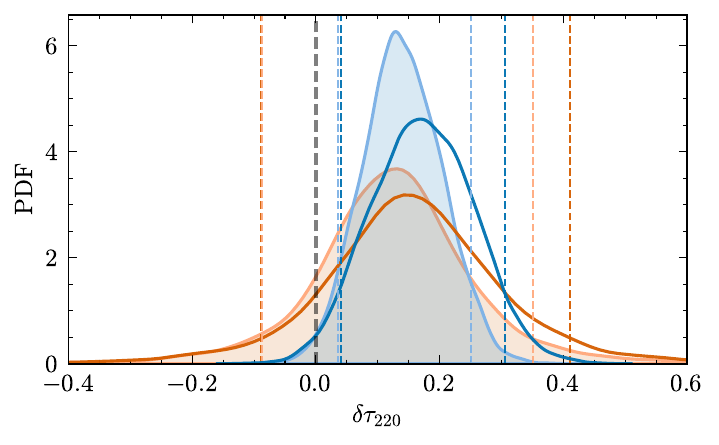}
    \caption{
        The one-dimensional combined constraints on $\df{220}$ and $\dtau{220}$, obtained by multiplying the posteriors from individual events (in blue), and using hierarchical combination (in orange), from GWTC-3 events. We compare results using the \texttt{pSEOBNRv5PHM} model (dark, unfilled curves) and the \texttt{pSEOBNRv4HM} model (light, filled curves). Vertical dashed lines represent the 90\% credible intervals.
    }
    \label{fig:combined_bound_comparison}
\end{figure*}

Compared to the result of Ref.~\cite{LIGOScientific:2021sio} for GW200129\_065458, we note that the posterior of $\delta \tau_{220}$ exhibits a larger tail towards positive values. This shift in $\delta \tau_{220}$ is driven by a slightly different recovery of the binary's luminosity distance and inclination angle when spin precession is included in the analysis.
When analyzing the event with \texttt{pSEOBNRv5HM} under the aligned-spin assumption, we find results consistent with Ref.~\cite{LIGOScientific:2021sio}, indicating that the difference arises solely from the inclusion of spin precession.
Overall, the \texttt{pSEOBNRv5PHM} model provides a significantly better fit to the data compared to \texttt{pSEOBNRv4HM}, with a natural log Bayes factor of $\ln \mathcal{B} \simeq 5.1$, favoring a spin-precessing quasicircular hypothesis over an aligned-spin quasicircular one.
Under the assumption of a spin-precessing quasicircular binary, the impact of waveform systematics for this event remains non-negligible. A notable feature absent in \texttt{pSEOBNRv5PHM} is the inclusion of multipole asymmetries in the coprecessing frame, which are important for capturing evidence of spin precession in this event~\cite{Kolitsidou:2024vub}. It would be valuable to reanalyze this event using future versions of \texttt{pSEOBNRv5PHM} that incorporate such asymmetries, as well as using a parametrized version of the recently developed aligned-spin, eccentric \texttt{SEOBNRv5EHM} model~\cite{Gamboa:2024hli, Gamboa:2024imd}, to further explore the impact of waveform assumptions on the results. We leave these studies for future work.

The posterior probability distributions for the fractional deviations in the frequency and damping time of the $(2,2,0)$ QNM ($\delta f_{2 2 0}$ and $\delta \tau_{2 2 0}$) for all 10 events from GWTC-3 used to produce combined results are reported in Appendix~\ref{sec:gwtc-3}. We specifically highlight the differences between the \texttt{pSEOBNRv5PHM} and \texttt{pSEOBNRv4HM} analyses.
Overall, the two models produce broadly consistent results across all events. For several events, tighter constraints are achieved on $\delta f_{2 2 0}$ with the \texttt{pSEOBNRv5PHM} model, particularly for GW190519\_153544, which no longer shows a secondary mode in the posterior distribution for $\delta f_{2 2 0}$ that was present in the \texttt{pSEOBNRv4HM} results. Results for $\delta \tau_{2 2 0}$ are also largely consistent, but a tail toward large $\delta \tau_{2 2 0}$ values is seen for GW200129\_065458, as previously noted, and is even more pronounced for GW200311\_115853.

Also for GW200311\_115853, this tail is driven by a different recovery of the binary's distance and inclination angle when spin precession is included in the analysis, while using \texttt{pSEOBNRv5HM} under the aligned-spin assumption gives results consistent with Ref.~\cite{LIGOScientific:2021sio}.
The posteriors for the distance and inclination recovered using \texttt{pSEOBNRv5HM} are also in agreement with those obtained using \texttt{SEOBNRv5PHM} assuming GR.
The maximum likelihood point recovered by \texttt{pSEOBNRv5PHM} corresponds to a positive $\delta \tau_{2 2 0}$, indicating a genuine correlation that provides a good fit to the data.
This is confirmed not to be a sampling issue, as additional runs with more stringent sampler settings ($\mathrm{nlive}=2000$, $\mathrm{maxmcmc}=10000$) yield consistent results. Furthermore, the natural log Bayes factor of $\ln \mathcal{B} \simeq 1.7$ favors \texttt{pSEOBNRv5PHM} under the spin-precessing hypothesis over the aligned-spin scenario.
We further investigate correlations between $\chi_{\mathrm{p}}$, $d_L$, $\iota$, and QNM deviation parameters in Appendix~\ref{sec:correlations}.

The combined bounds on the fractional deviations in the frequency and damping time of the $(2,2,0)$ QNM using the \texttt{pSEOBNRv5PHM} model read
\begin{equation}
    \delta {f}_{220}=0.01_{-0.04}^{+0.04}
    \quad \textrm{and} \quad
    \delta {\tau}_{220}=0.17_{-0.13}^{+0.14},
\end{equation}
by multiplying the posteriors and
\begin{equation}
    \begin{aligned}
    & \delta {f}_{220}=0.00_{-0.06}^{+0.06} \quad\left[\mu=0.00_{-0.03}^{+0.03},\,\,  \sigma<0.05\right] \\
    & \delta {\tau}_{220}=0.15_{-0.24}^{+0.26} \quad\left[\mu=0.15_{-0.15}^{+0.15},\,\, \sigma<0.22\right]
    \end{aligned}
\end{equation}
by combining hierarchically. The numbers in the square brackets are the hyper-parameter estimates.
These results are broadly consistent with those reported by the LVK Collaboration from the analysis of the same 10 GW events from GWTC-3~\cite{LIGOScientific:2021sio}, which used the \texttt{pSEOBNRv4HM} model without accounting for spin-precession effects.

In Fig.~\ref{fig:combined_bound_comparison} we compare the combined constraints on $\df{220}$ and $\dtau{220}$ from GWTC-3 events, obtained using the \texttt{pSEOBNRv5PHM} model (dark, unfilled curves) and the \texttt{pSEOBNRv4HM} model (light, filled curves).
Constraints obtained by multiplying the posteriors from individual events are shown in blue, while results using hierarchical combination are shown in orange. Vertical dashed lines indicate the 90\% credible intervals. The gray vertical dashed lines mark the GR predictions $(\delta f_{220}, \delta \tau_{220})=(0,0)$

The updated analysis provides slightly tighter constraints on $\delta f_{220}$ compared to previous results, particularly when combining events hierarchically. This improvement is consistent with narrower posteriors obtained for several single-event results. Additionally, the peak of the posterior for $\delta f_{220}$ is closer to zero in both the joint and hierarchically combined analyses.
For $\delta \tau_{220}$, both the joint and hierarchically combined posteriors are broader and slightly shifted towards positive values. This is consistent with the presence of tails in single-event posteriors for specific events such as GW200311\_115853 and GW200129\_065458.

Similar to the results from GWTC-3, the joint posterior distribution for $\delta \tau_{220}$ places the GR prediction near the edge of the 90\% credible level. This discrepancy could arise from a variety of factors, including noise fluctuations~\cite{Ghosh:2021mrv, LIGOScientific:2020tif}, parameter correlations~\cite{LIGOScientific:2021sio}, or intrinsic variance due to the limited number of events in the catalog~\cite{Pacilio:2023uef}. Incorporating additional events from ongoing observing runs could help clarify this behavior.

Given a two-dimensional posterior distribution $P(x,y)$, the consistency with the null hypothesis can also be quantified by the GR quantile,
\begin{equation}
    Q_0 = \int_{\substack{\{x, y\} \text{ where } \\ P(x, y) \geq P\left(0, 0\right)}} P(x, y) \, \mathrm{d} x \, \mathrm{d} y.
\end{equation}
The GR quantile corresponds to the fraction of the posterior enclosed by the isoprobability contour that passes through the GR value $(0, 0)$~\cite{Ghosh:2017gfp}, and is defined such that $Q_0 = 0$ ($Q_0=1$) indicates full consistency (full inconsistency) with the null hypothesis.
For the joint posteriors, we consider $\{x, y\} = \{\delta f_{220}, \delta \tau_{220}\}$, while for the hierarchically results we have $\{x, y\} = \{\mu, \sigma^2\}$.
We summarize the GR quantiles in Table~\ref{tab:gr_quantiles}.
Using the \texttt{pSEOBNRv5PHM} model, we find $Q_0=0.93$ when multiplying the posteriors, while the hierarchical combination yields $Q_0=0.003$ for $\delta f_{220}$ and $Q_0=0.74$ for $\delta \tau_{220}$.
For the \texttt{pSEOBNRv4HM} model, the GR quantile when multiplying the posteriors is $Q_0=0.97$, whereas the hierarchical combination gives $Q_0=0.32$ for $\delta f_{220}$ and $Q_0=0.81$ for $\delta \tau_{220}$.
Overall, the \texttt{pSEOBNRv5PHM} model shows slightly better consistency with GR in all cases.

The \texttt{pSEOBNRv5PHM} model can also be used to estimate the properties of the remnant BH. We compute effective values for the QNM frequency and damping time as follows:
\begin{subequations}
\begin{align}
    f_{\ell m 0} & =f_{\ell m 0}^{\mathrm{GR}}\left(1+\delta {f}_{\ell m 0}\right), \\
    \tau_{\ell m 0} & =\tau_{\ell m 0}^{\mathrm{GR}}\left(1+\delta {\tau}_{\ell m 0}\right),
\end{align}
\end{subequations}
where $f_{\ell m 0}^{\mathrm{GR}}$ and $\tau_{\ell m 0}^{\mathrm{GR}}$ are derived as functions of the component masses and spins using NR fits. The mass and spin of the remnant object can then be estimated from the complex QNM frequencies by inverting the fitting formula provided in Ref.~\cite{Berti:2005ys}. These results are summarized in Table~\ref{tab:qnm_o1o2o3_results}.

\begin{table}[t]
    \centering
    \renewcommand{\arraystretch}{1.2}
    \setlength{\tabcolsep}{6pt}
    \begin{tabular}{lccc}
        \hline
        Model & Joint \(Q_0\) & \multicolumn{2}{c}{Hierarchical \(Q_0\)} \\
              &              & \(\delta f_{220}\) & \(\delta \tau_{220}\) \\
        \hline
        \texttt{pSEOBNRv4HM}  & 0.97 & 0.32 & 0.81 \\
        \texttt{pSEOBNRv5PHM}  & 0.93 & 0.003 & 0.74 \\
        \hline
    \end{tabular}
    \caption{Comparison of GR quantiles (\(Q_0\)) for joint and hierarchical results using \texttt{pSEOBNRv4HM} and \texttt{pSEOBNRv5PHM}.
    Smaller values of \(Q_0\) indicate better consistency with GR. }
    \label{tab:gr_quantiles}
\end{table}



\section{Conclusions}
\label{sec:conclusions}

We have presented \pseob, a parametrized, multipolar, waveform model for BBHs in quasicircular orbits, designed to perform null tests of GR across the inspiral, plunge-merger, and ringdown stages of compact binary coalescences.
Notably, \pseob{} extends previous works which were limited to BBHs with aligned or antialigned spins~\cite{Ghosh:2021mrv, Maggio:2022hre, Toubiana:2023cwr} by incorporating spin-precession effects.
After examining the morphology of the parametrized deviations, we employed \pseob{} to estimate deviations in the QNM frequency and damping time of the $(\ell,m,n) = (2,2,0)$ mode. Synthetic-signal studies using BBH NR waveforms highlight the importance of including spin-precession effects, even at current detector sensitivities, to robustly perform tests of GR and avoid biases or false deviations.
By analyzing a synthetic signal from a publicly available NR simulation of scalar-field BS merger, we showed that \pseob{} can successfully identify that the signal does not originate from a BBH in GR.
Finally, we have applied our model to real data by reanalyzing 12 events from GWTC-3. Using a hierarchical combination of these events, we constrained fractional deviations in the frequency and damping time of the $(2,2,0)$ quasinormal-mode to $\delta {f}_{220}=0.00_{-0.06}^{+0.06}$ and $\delta {\tau}_{220}=0.15_{-0.24}^{+0.26}$ at 90\% credibility.
Our results are consistent with those from the LVK Collaboration, which did not account for spin-precession effects.

Similar to the results from GWTC-3, the joint posterior distribution for $\delta \tau_{220}$ places the GR prediction near the edge of the 90\% credible level. Analyzing additional events from current and upcoming LVK observing runs is expected to refine these constraints and help better quantify the significance of the results.
Furthermore, performing large-scale injection studies using NR or NR surrogate waveforms, along with incorporating Gaussian noise realizations or real-noise injections, would offer important insights into the potential influence of waveform systematics or noise fluctuations on the inferred deviation parameters. These effects should be thoroughly investigated and quantified in order to be able to claim a GR violation in GW observations~\cite{Gupta:2024gun}.

In this work, while we described the implementation and morphology of all corrections, we validated our model through Bayesian PE for the ringdown of the dominant $(\ell,m,n)=(2,2,0)$ mode, as it is the primary test currently performed in \LVK{} analyses. At current \snr{}, higher modes in the ringdown are not detectable with high statistical significance, but are expected to become confidently detectable with the improved \snr{} achievable through upcoming LVK upgrades~\cite{Brito:2018rfr, Gennari:2023gmx}. In future work, we plan to explore higher modes in the ringdown, as well as the measurability of deviations in the inspiral and plunge-merger stages.

Given the modular nature of the \texttt{pySEOBNR} code, ongoing improvements to the baseline \texttt{SEOBNRv5PHM} model, such as the inclusion of multipole asymmetries in the coprecessing frame, and the calibration to spin-precessing NR simulations, can be immediately propagated to the parametrized \pseob{} model. It is also straightforward to add similar parametrized deviations to the eccentric, aligned-spin \texttt{SEOBNRv5EHM} waveform model~\cite{Gamboa:2024hli, Gamboa:2024imd} recently developed.

Massive BH binaries with masses from $10^4\,M_{\odot}$ to $10^7\,M_{\odot}$, detectable by the space-based LISA mission, are also prime candidates for BH spectroscopy tests~\cite{Berti:2005ys}. The high SNR of these sources makes it especially important to incorporate all relevant physical effects in their analysis, including spin-precession, to mitigate potential biases arising from waveform systematics. We plan to use the \texttt{pSEOBNRv5PHM} model to extend the work of Ref.~\cite{Toubiana:2023cwr} to spin-precessing binaries, also including a realistic treatment of the LISA response function.


\section*{Acknowledgements}

\label{sec:acknowledgements}
%
%
We are grateful to Gregorio Carullo, Eleanor Hamilton, Danny Laghi, Sylvain Marsat, and Manuel Piarulli for performing the LIGO-Virgo-KAGRA review of the \texttt{pSEOBNRv5PHM} model. We also thank Raffi Enficiaud for his assistance during the review and for valuable discussions, Yotam Sherf for collaboration during the early stages of this project, and Tamara Evstafyeva for providing additional details on the boson-star simulations. We thank Juan Calderon Bustillo, Neil Lu, and Harrison Siegel for their helpful comments on the manuscript.

%
E.M. and H.O.S acknowledge funding by the Deutsche Forschungsgemeinschaft (DFG)~-~project No.:~386119226.
E.M. is supported by the European Union's Horizon Europe research and innovation
programme under the Marie Skłodowska-Curie grant agreement No.~101107586.
%
%
We also acknowledge the computational resources provided by the Max Planck Institute
for Gravitational Physics (Albert Einstein Institute), Potsdam, in particular,
the \texttt{Hypatia} cluster.

%
%
The material presented in this paper is based upon work supported by National
Science Foundation's (NSF) LIGO Laboratory, which is a major facility fully
funded by the NSF.
%
%
This research has made use of data or software obtained from the Gravitational
Wave Open Science Center (\href{gwosc.org}{gwosc.org}), a service of LIGO
Laboratory, the LIGO Scientific Collaboration, the Virgo Collaboration, and
KAGRA. LIGO Laboratory and Advanced LIGO are funded by the United States
National Science Foundation (NSF) as well as the Science and Technology
Facilities Council (STFC) of the United Kingdom, the Max-Planck-Society (MPS),
and the State of Niedersachsen/Germany for support of the construction of
Advanced LIGO and construction and operation of the GEO600 detector.
Additional support for Advanced LIGO was provided by the Australian Research
Council. Virgo is funded, through the European Gravitational Observatory (EGO),
by the French Centre National de Recherche Scientifique (CNRS), the Italian
Istituto Nazionale di Fisica Nucleare (INFN) and the Dutch Nikhef, with
contributions by institutions from Belgium, Germany, Greece, Hungary, Ireland,
Japan, Monaco, Poland, Portugal, Spain. KAGRA is supported by Ministry of
Education, Culture, Sports, Science and Technology (MEXT), Japan Society for
the Promotion of Science (JSPS) in Japan; National Research Foundation (NRF)
and Ministry of Science and ICT (MSIT) in Korea; Academia Sinica (AS) and
National Science and Technology Council (NSTC) in
Taiwan.

\section*{Data availability}

The data that support the findings of this article are openly available~\cite{KAGRA:2023pio}.

\appendix


\section{Posterior distributions of GWTC-3 events}
\label{sec:gwtc-3}

\begin{figure*}
    \includegraphics[width=0.3\textwidth]{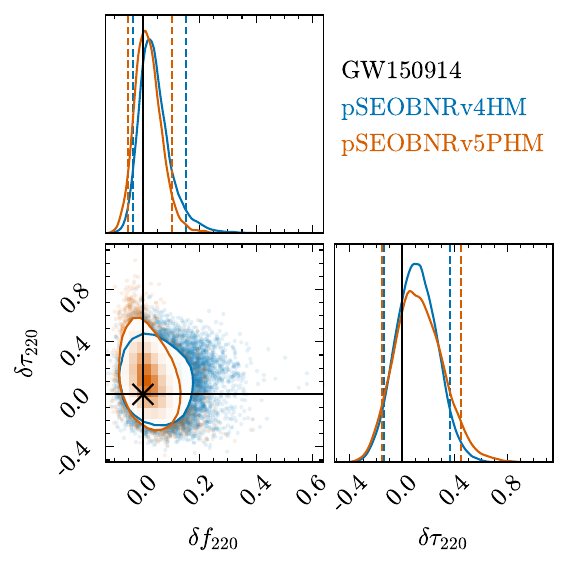}
    \includegraphics[width=0.3\textwidth]{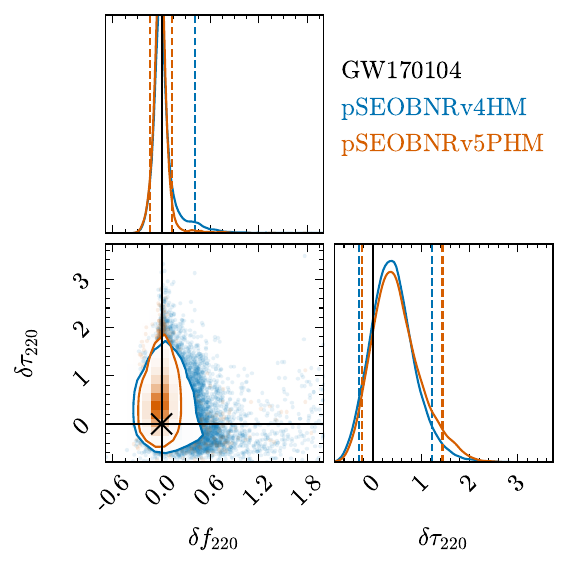}
    \includegraphics[width=0.3\textwidth]{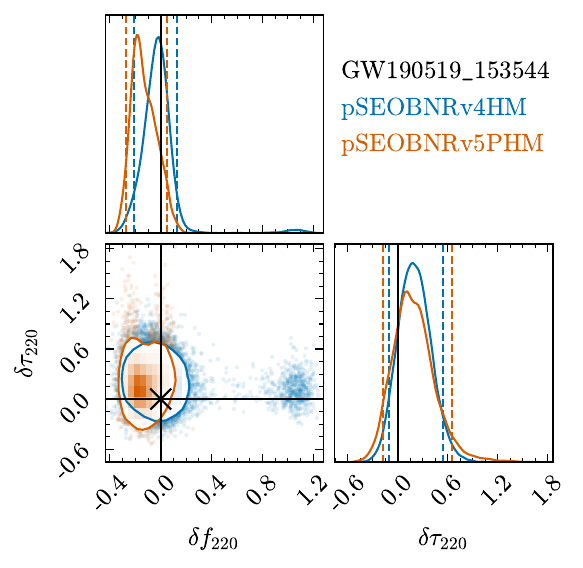}
    \includegraphics[width=0.3\textwidth]{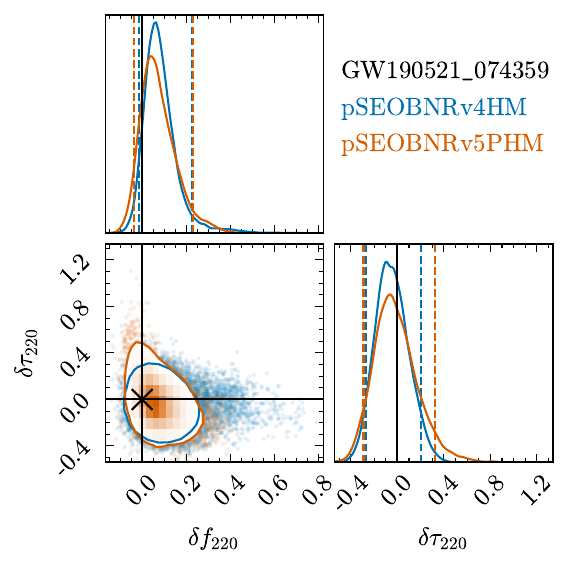}
    \includegraphics[width=0.3\textwidth]{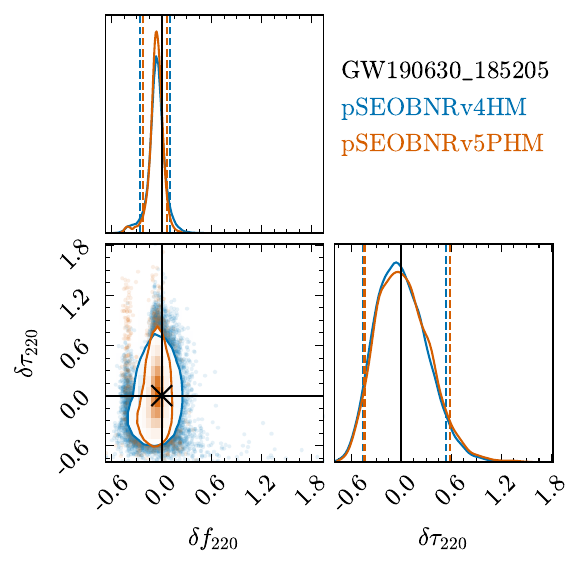}
    \includegraphics[width=0.3\textwidth]{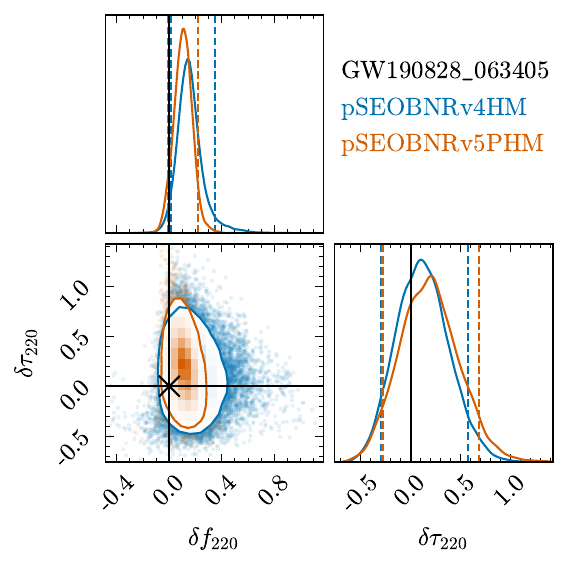}
    \includegraphics[width=0.3\textwidth]{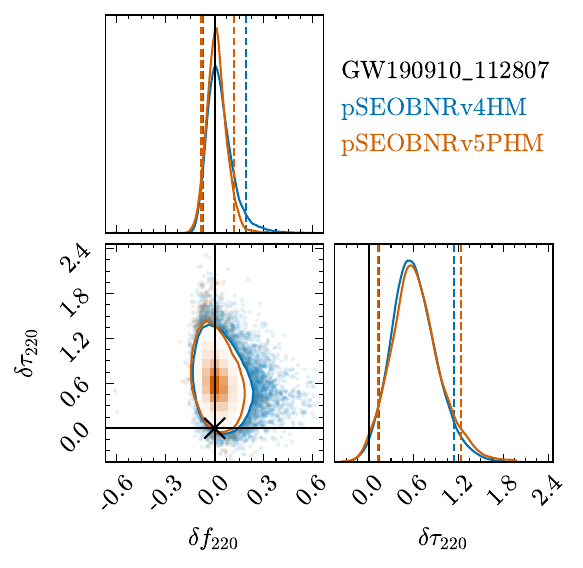}
    \includegraphics[width=0.3\textwidth]{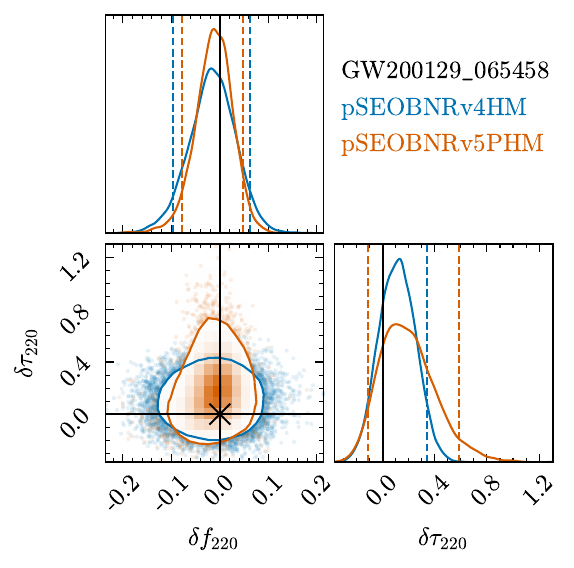}
    \includegraphics[width=0.3\textwidth]{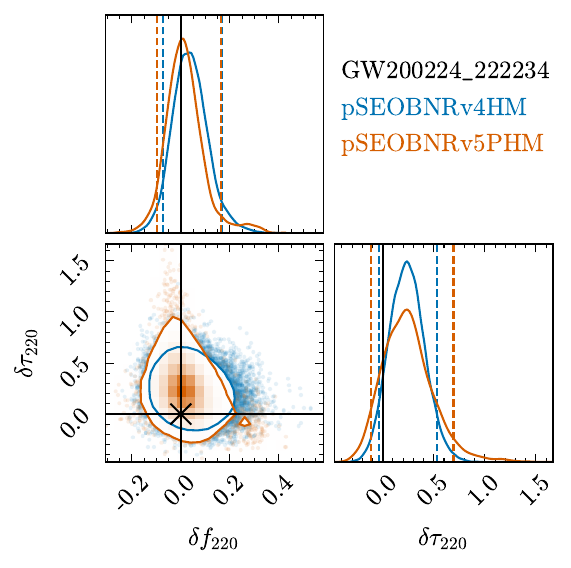}
    \includegraphics[width=0.3\textwidth]{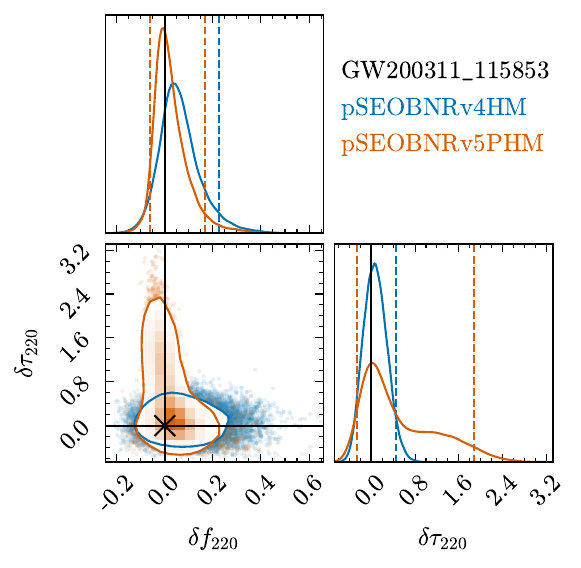}
    \caption{Posterior probability distributions for the fractional deviations in the frequency and damping time of the $(2,2,0)$ QNM ($\delta f_{2 2 0}$ and $\delta \tau_{2 2 0}$), for the 10 events from GWTC-3 used to produce combined results for the \texttt{pSEOBNR} analysis.
    The PE is performed with the \texttt{pSEOBNRv4HM} model (blue) and with the \texttt{pSEOBNRv5PHM} model (orange).
    The 2D contours mark the 90\% credible regions, while the dashed lines on the 1D marginalized distributions mark the 90\% credible intervals. The black vertical and horizontal lines mark the GR predictions ($\delta f_{2 2 0} = \delta \tau_{2 2 0} = 0$).
    }
    \label{fig:corner-gwtc-3}
\end{figure*}

We show in Fig.~\ref{fig:corner-gwtc-3} the posterior probability distributions for the fractional deviations in the frequency and damping time of the $(2,2,0)$ QNM ($\delta f_{2 2 0}$, $\delta \tau_{2 2 0}$), for the 10 events from GWTC-3 used to produce combined results for the \texttt{pSEOBNR} analysis.
The PE is performed with the \texttt{pSEOBNRv4HM} model (blue) and with the \texttt{pSEOBNRv5PHM} model (orange). The results from \texttt{pSEOBNRv4HM} are taken from the data release associated with Ref.~\cite{LIGOScientific:2021sio}.
The 2D contours mark the 90\% credible regions, while the dashed lines on the 1D marginalized distributions mark the 90\% credible intervals. The black vertical and horizontal lines mark the GR predictions, $\delta f_{2 2 0} = \delta \tau_{2 2 0} = 0$.

\section{Correlations between GR parameters and QNM deviations for GW200311\_115853}
\label{sec:correlations}
\begin{figure*}
    \includegraphics[width=0.8\textwidth]{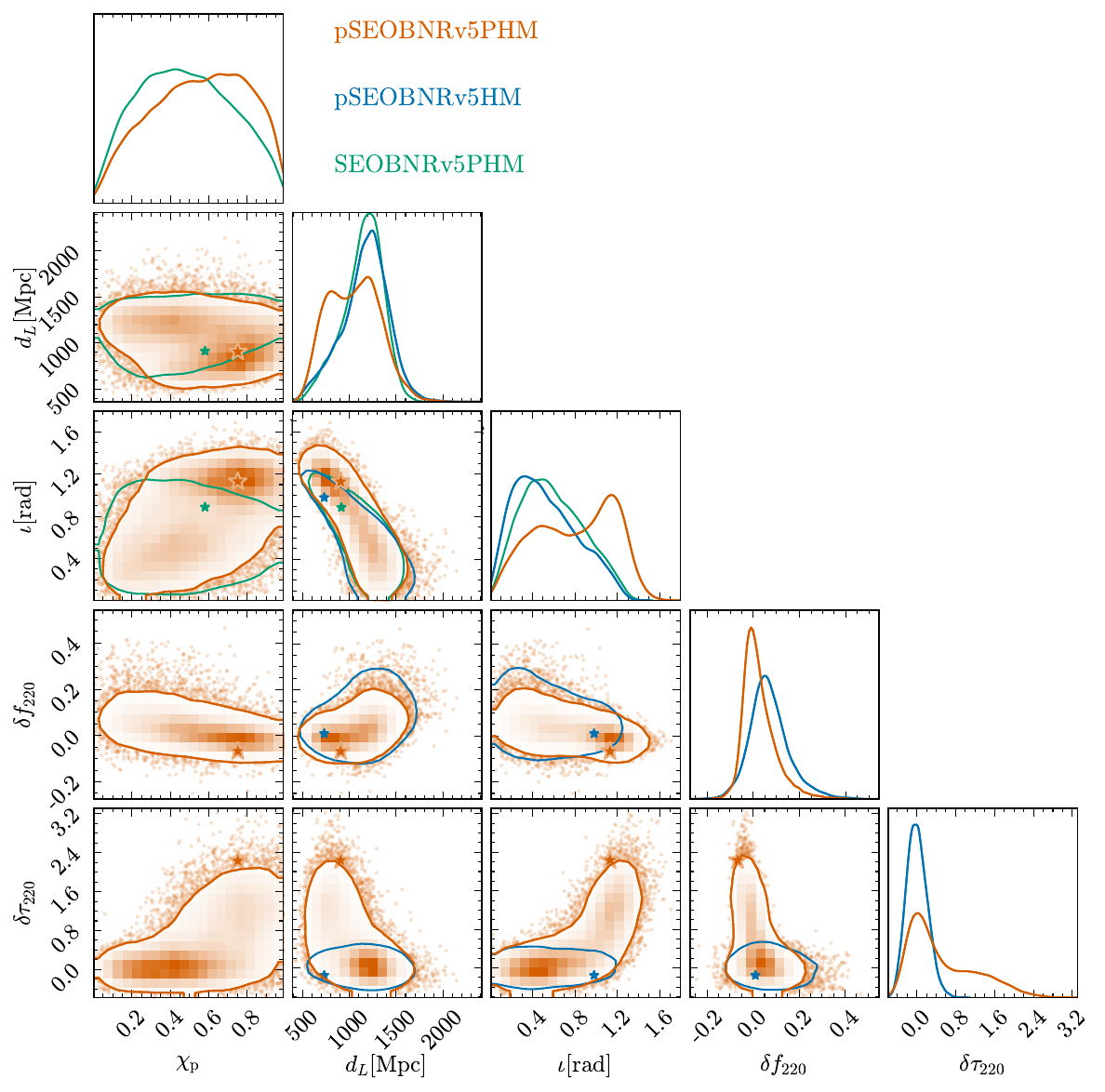}
    \caption{Posterior probability distributions for the effective precessing-spin parameter $\chi_{\mathrm{p}}$, the binary's luminosity distance $d_L$ and inclination angle $\iota$, and the fractional deviations in the frequency and damping time of the $(2,2,0)$ QNM ($\delta f_{2 2 0}$, $\delta \tau_{2 2 0}$), for the event GW200311\_115853. We show recoveries using the \texttt{pSEOBNRv5PHM} model (in orange) and its aligned-spin version \texttt{pSEOBNRv5HM} (in blue), as well the parameters inferred assuming GR with the \texttt{SEOBNRv5PHM} model (in green). The 2D contours mark the 90\% credible regions, and the stars indicate the maximum likelihood parameters recovered in each analysis.}
    \label{fig:correlations}
\end{figure*}

As mentioned in Sec.~\ref{sec:pe_real_data} of the main text, for GW200311\_115853 we observe a tail in the posterior distribution for $\delta \tau_{220}$ towards positive values, when spin precession is included.
Figure~\ref{fig:correlations} shows the posterior probability distributions for the fractional deviations in the frequency and damping time of the $(2,2,0)$ QNM ($\delta f_{2 2 0}$, $\delta \tau_{2 2 0}$), along with the effective precessing-spin parameter $\chi_{\mathrm{p}}$, the binary's luminosity distance $d_L$, and inclination angle $\iota$, highlighting the correlations among these parameters. The 2D contours mark the 90\% credible regions. We show recoveries using the \texttt{pSEOBNRv5PHM} model (in orange) and its aligned-spin version \texttt{pSEOBNRv5HM} (in blue), as well the parameters inferred assuming GR with the \texttt{SEOBNRv5PHM} model (in green).
The stars indicate the maximum likelihood parameters in each analysis.

Comparing the \texttt{pSEOBNRv5PHM} and \texttt{pSEOBNRv5HM} recoveries, we note that correlations between $\delta \tau_{2 2 0}$ and ($d_L$, $\iota$) appear only when spin precession is included.
In particular, the tail toward smaller distances and higher inclinations is correlated with $\chi_{\mathrm{p}}$, which in turn is correlated with $\delta \tau_{2 2 0}$.
However, the \texttt{SEOBNRv5PHM} results show that correlations between ($d_L$, $\iota$) and $\chi_{\mathrm{p}}$ are not as pronounced in the GR recovery.
When examining the two-dimensional posterior for distance and inclination, the recoveries from \texttt{SEOBNRv5PHM} and \texttt{pSEOBNRv5HM} are broadly consistent, while \texttt{pSEOBNRv5PHM} exhibits a secondary mode at smaller distances and higher inclinations.
Interestingly, the maximum likelihood for both \texttt{SEOBNRv5PHM} and \texttt{pSEOBNRv5HM} lies near the secondary mode of \texttt{pSEOBNRv5PHM}, despite most of the posteriors being centered elsewhere. This high likelihood indicates that even the \texttt{SEOBNRv5PHM} and \texttt{pSEOBNRv5HM} models fit the data well in that region; the increased flexibility provided by including both spin precession and QNM deviations allows for a broader range of configurations that match well the data around the region, at the cost of a shift in $\delta \tau_{2 2 0}$.

\bibliography{biblio}

\end{document}